\documentclass{article}
\usepackage{arxiv}
\usepackage[utf8]{inputenc} 
\usepackage[T1]{fontenc}    
\usepackage{hyperref}       
\usepackage{url}            
\usepackage{booktabs}       
\usepackage{amsfonts}       
\usepackage{nicefrac}       
\usepackage{microtype}      
\usepackage{lipsum}
\usepackage{graphicx}

\usepackage{bm}
\usepackage{amsmath}
\usepackage{relsize}
\usepackage{rotating} 
\usepackage{multirow}

\title{Optimal Experimental Design for Mathematical Models of Hematopoiesis}

\author{
  Luis Martinez Lomeli\thanks{These authors contributed equally to this work} \\
  Center for Complex Biological Systems\\
  UC Irvine, CA, USA\\
   \And
  Abdon Iniguez\footnotemark[1] \\
  Center for Complex Biological Systems\\
  UC Irvine, CA, USA\\
  \And
  Prisanthi Tata  \\
  Division of Hematology/Oncology\\
  UC Irvine, CA, USA\\
  \And
  Nilamani Jena  \\
  Division of Hematology/Oncology\\
  UC Irvine, CA, USA\\
  \And
  Zhong-Ying Liu  \\
  Division of Hematology/Oncology\\
  UC Irvine, CA, USA\\
  \And
  Richard Van Etten  \\
  Division of Hematology/Oncology\\
  Department of Biological Chemistry\\
  Center for Cancer Systems Biology\\
  Center for Complex Biological Systems\\
  Chao Family Comprehensive Cancer Center\\
  UC Irvine, CA, USA\\
  \And
  Arthur D. Lander  \\
  Department of Developmental and Cell Biology\\
  Department of Biomedical Engineering\\
  Center for Cancer Systems Biology\\
  Center for Complex Biological Systems\\
  Chao Family Comprehensive Cancer Center\\
  UC Irvine, CA, USA\\
  \And
  Babak Shahbaba\thanks{Corresponding authors: babaks@uci.edu, jlowengr@uci.edu, vminin.uci.edu} \\
  Department of Statistics\\
  Center for Cancer Systems Biology\\
  Center for Complex Biological Systems\\
  UC Irvine, CA, USA\\
  \And
  John S. Lowengrub\footnotemark[2]\\
  Department of Mathematics\\
  Department of Biomedical Engineering\\
  Center for Cancer Systems Biology\\
  Center for Complex Biological Systems\\
  Chao Family Comprehensive Cancer Center\\
  UC Irvine, CA, USA\\
  \And
  Vladimir N. Minin\footnotemark[2]\\
  Department of Statistics\\
  Center for Cancer Systems Biology\\
  Center for Complex Biological Systems\\
  UC Irvine, CA, USA\\
}

\begin{document}
\maketitle
\begin{abstract}
The hematopoietic system has a highly regulated and complex structure in which cells are organized to successfully create and maintain new blood cells.
It is known that feedback regulation is crucial to tightly control this system, but the specific mechanisms by which control is exerted are not completely understood. 
In this work, we aim to uncover the underlying mechanisms in hematopoiesis by conducting perturbation experiments, where animal subjects are exposed to an external agent in order to observe the system response and evolution. We have developed a novel Bayesian framework for optimal design of perturbation experiments and proper analysis of data collected from these experiments. 
Here, we consider an experiment, where mice are exposed to a low dose of radiation, which reduces the number of hematopoietic stem cells but leaves the progenitor and other hematopoietic cells in the bone marrow largely unchanged.
We use a mechanistic differential equation model that accounts for feedback and feedforward regulation on cell division rates and self-renewal probabilities. 
A significant obstacle is that the experimental data are not longitudinal, rather each data point corresponds to a different animal. We overcome this difficulty by developing a hierarchical Bayesian framework with latent variables that capture unobserved cellular population levels. We then use principles of Bayesian experimental design to optimally distribute time points at which the numbers of bone marrow cells are observed. We evaluate our approach using synthetic data and real experimental data and show that an optimal design can lead to better estimates of model parameters.

\end{abstract}


\section{Introduction}
The hematopoietic system produces billions of mature myeloid and lymphoid blood cells from self-renewing hematopoietic stem cells (HSCs) and multi-potent progenitors (MPPs) on a daily basis and facilitates massive cell increases in response to pathological stresses \cite{rieger2012hematopoiesis}.
This system must have in place a tightly regulated feedback control mechanism at multiple levels to ensure an appropriate proportion of HSCs, MPPs, and mature cells. 
However, we do not yet have a good understanding of the nature of the feedback regulation and how it plays a role in cell maintenance. 
In this work, we present a hierarchical  framework for modeling the hematopoiesis system with feedback control and regulation.
Our approach is based on a rigorous Bayesian methodology for fitting mechanistic mathematical models to empirical data. 
Using a utility-based theory, we provide a rigorous procedure towards the determination of an optimal design of experiment when the goal is to estimate the parameters of the hierarchical model.
A key challenge is that in the experiments used here, the data is not longitudinal as the method of data collection involves destruction of the source (e.g., sacrifice of a mouse).
Therefore, each data point corresponds to a different experiment.

There has been a longstanding effort to use mathematical models in order to understand hematopoiesis under normal and diseased conditions, e.g., see \cite{SysBioBlood,Hofer2016,Pujo-Menjouet2016,MacLean2017,Fornari2018}. These include ordinary differential equation (ODE) models that describe the dynamics of simplified systems (e.g., \cite{michor2005dynamics,Marciniak2009,busch2015fundamental,Craig2016,Glauche2018,Mahadik2019computational}), models that account for more realistic numbers of different cell types and branching processes \cite{manesso2013dynamical}, as well as models that account for stochasticity \cite{roeder2002,Dingli2007,Kimmel2014,Rozhik2016,Jakel2018,Xu2018} and spatial dynamics in the bone marrow \cite{krinner2013merging}. 
In many cases, models were fitted using equilibrium cell counts, or limited dynamic data, which yield point estimates for the parameters.  
In a few cases, uncertainties in parameter inference were considered using Bayesian methods \cite{golinelli2006bayesian,fong2009bayesian,xu2019statistical,oden2009toward, farrell2017adaptive}.

Here, we use a nonlinear ODE model that incorporates self-renewal, cell division, feedback and feedforward regulation for a simplified description of hematopoiesis. In particular, we track only stem and multipotent progenitor cells. Nevertheless, the model is flexible enough to describe the response to external perturbations and the subsequent return to steady state.

We apply our model to empirical data obtained from perturbation experiments  in mice subjected to low dose radiation. We measure the numbers of HSCs and MPPs in the bone marrow to investigate the recovery dynamics and infer model parameters and feedback mechanisms.
Note that the mice can not be tracked longitudinally by taking repeated measurements of cell numbers; rather, each mouse provides a single observation point because the mouse is sacrificed to extract the bone marrow.
This creates a statistical challenge for quantifying the system response since each data point belongs to a different subject, for whom we do not have the baseline measurements prior to exposure to the agent; that is, the number of cells is not known initially or at any other time prior to the measurement time and thus the corresponding cell numbers are latent.
To address this issue, we use a hierarchical Bayesian model where the initial cellular counts for all the subjects are treated as latent variables to be inferred  and the ODE model is used to interpolate the cell numbers until the observation times.
The main advantage of this approach is that it allows us to appropriately integrate and align data from multiple subjects in a coherent, statistically rigorous manner
Note that we still need to determine how sensitive our estimates are to the choice of experimental design.

Designing experiments for investigating the hematopoietic process typically involves specifying a set of variables such as: timing of the measurements, number of subjects per observation time, nature of the perturbation, etc. 
Our goal is to determine the optimal values of these variables to maximize the information gain for the parameters of our ODE model.
Similar approaches have been used in other areas \cite{dehideniya2018optimal, zhang2018optimal,han2004bayesian}. 
Using a Bayesian utility theory approach, we quantify information gain about the ODE parameters over the space of all possible experimental designs. 
More specifically, we use the Kullback-Leibler divergence \cite{kullback1997information,ryan2016review} to quantify the difference between the prior and posterior distributions of the parameters.
This way, we are able identify the  design that provides the highest expected utility, e.g., maximum information gain \cite{chaloner1995bayesian,lindley1956measure,cook2008optimal,huan2013simulation,muller1995optimal,wakefield1994expected,palmer1998bayesian, drovandi2013sequential,muller1999simulation, biegler2011large, muller2006bayesian, liepe2013maximizing, silk2014model}, which we call the optimal design.

To evaluate this approach, we first apply it to synthetic data and show that we can identify the ODE model parameters. 
Next, we analyze real data from a bone marrow perturbation experiment. 
Investigating a finite set of experimental designs, we find that the designs with higher number of observation times and possibly fewer subject replicates can provide better parameter estimates compared to the designs with fewer observation times even if we use a higher number of subject replicates.
Also, we show that how we allocate the subjects over time matters. 
For example, designs with more observations at later times can provide better estimates on feedback gains, whereas designs with more observations at earlier times improve identification of cell division rates.

\section{Materials and Methods\label{sec:Methods}}
\paragraph{Experimental setup}
We consider an experimental set up where the hematopoietic system is perturbed and the results of the perturbation are observed by measuring the numbers of the cell types of interest. 
We primarily consider hematopoietic stem cells (HSCs) and multipotent progenitors (MPPs), but other known cell types like lymphoid and myeloid progenitors  (CLPs, CMPs) or mature lymphoid and myeloid cells can also be quantified experimentally.
Throughout this paper, we will use simulated and real data based on the following experiment. 
We start with $M$ genetically identical mice that are kept under the same laboratory conditions. 
Each mouse is exposed to an external perturbation, e.g. a light dose (50 cGy) of radiation, with the purpose of decreasing the number of HSCs  (e.g., \cite{quesenberry1998}).
These mice are sacrificed at different times after irradiation and the counts of HSCs and MPPs are obtained from the bone marrow of each individual mice using flow cytometry. 
More information about the experimental materials and methods can be found in the Supplemental Materials.

\paragraph{Data generation process} 
We postulate that at time $t_0$, HSC and MPP counts come from some distribution that encapsulates normal biological variation among mice.
For mathematical convenience, we assume that this distribution is lognormal, or equivalently, that log-transformed unobserved true HSC and MMP counts come from two independent normal distributions with means $\log(\mu_{\text{HSC}})$ and $\log(\mu_{\text{MPP}})$ respectively, with the same variance $\sigma_b$.  
Denoting these latent log-transformed counts with a bivariate vector $\mathbf{u}_i$ and their means with a vector $\bm{\mu}$ for each mouse $i = 1,\dots,M$, we can write our initial condition assumption as

\begin{equation}
    \mathbf{u}_{i}\sim  N\left(\log (\bm{\mu}), \sigma_b^2\cdot \mathbf{I}\right), i=1,\dots,M. 
    \label{eq:likelihoodlatentvars}
\end{equation}
We are now ready to specify the distribution of observed HSC and MPP counts.
First, we order mice in such a way that mice indexed by $1,\dots,k$ correspond to animals, whose bone marrow is sampled immediately post perturbation.
We assume that conditionally on the true log-counts of HSCs and MPPs, the observed log-transformed counts $\mathbf{y}_i = \log(\mathbf{y}_i^*)$, where $\mathbf{y}_i^*$ represents the raw counts, are normally distributed, with the mean being equal to the true counts and variance $\sigma_t ^2$ that represents technical variation that arises due to the measurement error/noise.

For mice that are sacrificed right after the perturbation experiment, this assumption translates into
the following conditional distribution:
\begin{equation}
    \begin{array}{rl}
        
        \mathbf{y}_i(t_0) \mid \mathbf{u}_i \sim & N\left(\mathbf{u}_i, \sigma_t^2\cdot \mathbf{I} \right), i=1,\dots,k. \\
    \end{array}
    \label{eq:likelihoodt0Cond}
\end{equation}

The unconditional distribution for the HSC and MPP counts at the initial time $t_0$ is derived from equations (\ref{eq:likelihoodlatentvars}) and (\ref{eq:likelihoodt0Cond}) as 

\begin{equation}
    \begin{array}{rl}
        \mathbf{y}_i(t_0) \sim & N(\log(\bm{\mu}),(\sigma_{t}^{2}+\sigma_{b}^{2})\cdot \mathbf{I}), i = 1,\dots, k.
    \end{array}
    \label{eq:likelihoodt0}
\end{equation}

\begin{figure}[t]
\centering
\includegraphics[width=0.95\textwidth]{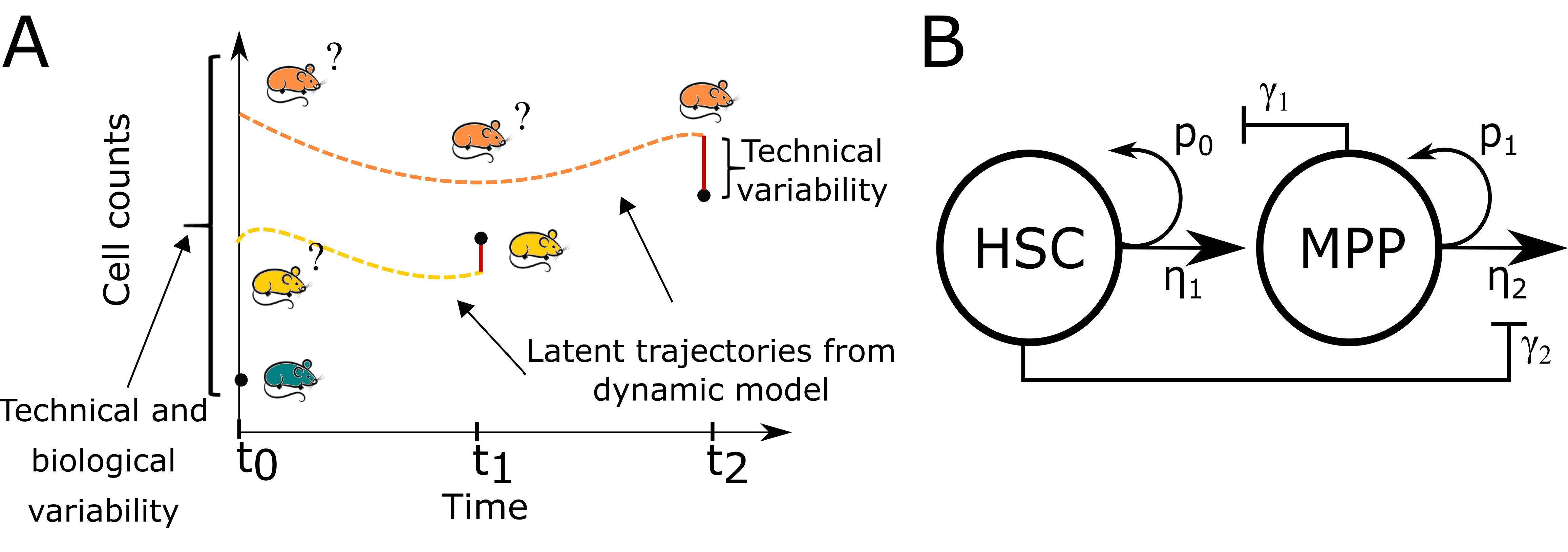}
\caption{\textbf{Illustration of the proposed latent variables approach and mechanistic model.} \textbf{A}. Description of the proposed latent variables approach. Each  mouse' cell counts are observed only once at the time of the mouse sacrifice and bone marrow extraction. Cell counts before perturbation (e.g., low dose radiation) are allowed to be different among mice due to normal biological variation. 
We model this by assuming each data point at at time $t_0$ to be subject to technical and biological variability. At times greater than $t_0$, we assume each data point has a latent trajectory subject to technical variability (shown in dashed). These latent trajectories are modeled using our mechanistic ODE model subject to these initial conditions.
\textbf{B}. The ODE lineage model consisting of HSC and MPP compartments. 
HSCs and MPPs have the ability to self renew with probabilities $p_0$ and $p_1$ and divide at rates$\eta_1$ and $\eta_2$). The HSCs self renewal probabilities are negatively regulated by the MPPs and the MPPs division rates are negatively regulated by the HSCs. See \cite{AbdonThesis} and text for details.}
  \label{fig:LineageODELatenVars}
  \vspace{-8pt}
\end{figure}

To model cell counts measured at time points after the initial time $t_0$,  we assume that the cellular population levels in each mouse start from latent initial conditions and follow deterministic latent trajectories according to a mechanistic process model. 
The latent population trajectories evolve for all times $0<t\leq t_j$ until the moment $t_j$ when the mouse is harvested and the cell counts are measured (with noise), see Figure \ref{fig:LineageODELatenVars}A. 
The conditional distribution of the observed HSC and MPP cell counts after the initial time given latent initial conditions is 

\begin{equation}
\begin{array}{c}
 \mathbf{y}_j(t_j) \mid \mathbf{u}_j
  \sim N(\log (\mathbf{x}(\mathbf{u}_j, \bm{\theta},t_j)),\sigma_{t}^{2} \cdot \mathbf{I}), j = k+1,\dots, M,
\end{array}
\label{eq:likelihoodt>0}
\end{equation}

where $\mathbf{x}(\mathbf{u}_j, \bm{\theta}, t_j)$ is a bivariate vector of HSC and MPP counts for mouse $j$ that started with latent counts $\mathbf{u}_j$ and evolved according to some process with parameters $\bm{\theta}$ up to time $t_j$. 
Note that equation (\ref{eq:likelihoodt>0}) does not include the biological noise term explicitly since it is already included in the model for initial cell counts.
This approach implies that the HSC and MPP trajectories are latent and can be observed cross-sectionally only once, in contrast to typical longitudinal studies, where repeated measures are taken from a cohort of animals/subjects followed over time.

\paragraph{Mechanistic model of the mean process} 
The dynamical model for latent trajectories is based on classic cell lineage models for describing the growth of hierarchically-organized tissues \cite{Marciniak2009,Lander2009,Buzi2015} where cells are arranged in a lineage starting with 
HSCs that are followed by more differentiated cells downstream. Although there are many cell types in the branched lineage that describes the hematopoietic  system (e.g., \cite{Dharampuriya2017,Brown2018}), we focus here only on the least differentiated types: the HSCs and MPPs that have the simple hierarchical relationship shown in Fig. \ref{fig:LineageODELatenVars}B. This is because the experiments suggest that the cell compartments downstream are largely unaffected by irradiation (data not shown) and thus we assume that the downstream cells do not significantly influence the HSC and MPP dynamics.  Further, for simplicity we do not distinguish between the different types of HSC and MPP cells, which eliminates the need for considering branching. Of course, the model can easily be extended to include more cell types and branching (e.g., \cite{krinner2013merging,manesso2013dynamical}) although this would require more data to constrain them.

In the mathematical model (Fig. \ref{fig:LineageODELatenVars}B), we assume that the HSCs and MPPs have the ability to divide at the rates $\tilde\eta_1^*$ and $\tilde\eta_2^*$, respectively, and to undergo self-renewal with probabilities $p_0^*$ and $p_1^*$. Let 
$x_{HSC}$ and $x_{MPP}$ be the numbers of HSCs and MPPs respectively, then their dynamics can be modeled using a system of ordinary differential equations: 
\begin{equation}
\begin{array}{l}
x'_{HSC} = (2p_0^*-1)\tilde\eta_1^* x_{HSC}, \\ \\
x'_{MPP} = 2(1-p_0^*)\tilde\eta_1^* x_{HSC} +(2p_1^*-1)\tilde\eta_2^* x_{MPP},\\ 
\end{array}
\label{eq:ODEreduced0}
\end{equation}
where $'=d/dt$.

The self-renewal probabilities and division rates should be subject to feedback regulation. Single cell RNA sequencing data (scRNA-seq) can be used to identify putative feedback loops and sender and receiver cells. We re-analyzed data from a scRNA-seq study of normal hematopoiesis that identified many interesting cell clusters whose transcriptomes suggested pairwise combinations of cells expressing feedback ligands and their receptors \cite{olsson2016single}. Although the study did not cleanly separate out different kinds of early stem/progenitor cells, the early stem/progenitors
did cluster into two groups, perhaps representing HSCs and MPPs. In these two groups, we
were able to recognize several ligands and receptors (such as ANGPT1 and CCL3 and their
receptors), although we could not be certain about the sender and receiver cell types. 

Following
\cite{Arai2004}, we hypothesized that ANGPT1 is secreted by HSCs and negatively regulates MPP division rates and that CCL3 is produced by MPPs and negatively regulates HSC self-renewal \cite{Staversky2018}. These hypotheses will be tested in future work. The feedback on HSC self-renewal can be modeled using a simple Hill function 
\begin{equation}
p_0^* = \frac{p_0}{1+\gamma_1 x_{MPP}},
\label{feedback on p0}
\end{equation}
where $p_0$ is the unregulated self-renewal probability and $\gamma_1$ is the feedback gain. Note that we have implicitly assumed that the concentration of the negatively regulating biomolecule (e.g.,  CCL3) is proportional to the cell population. This approximation assumes that spatial variation of the biomolecule can be neglected and is similar to the those used in \cite{Marciniak2009,Lander2009,manesso2013dynamical}. Although the HSC division rates could be subject to similar feedback as $p_0^*$ \cite{Lander2009}, we assume for simplicity that these rates are constant: $\tilde\eta_1^*=\eta_1$.

The negative feedforward loop on MPP division rates can be modeled as
 \begin{equation}
\tilde\eta_2^* = \frac{\tilde\eta_2}{1+\gamma_2 x_{HSC}},
 \label{feedback on eta2}
\end{equation}
where $\tilde \eta_2$ is the unregulated MPP division rate and $\gamma_2$ is the feedforward gain. 

In principle, the MPP self-renewal probability $p_1^*$ should also be subject to feedback regulation. Assuming this regulation arises from more differentiated cell types (e.g., \cite{Marciniak2009,Lander2009,manesso2013dynamical}), we can assume that $p_1^*=p_1$ is constant here since the more differentiated cells are largely unaffected by radiation.  Since the MPP should not be able to fully self-renew, $p_1<0.5$. Therefore, we may 
rewrite the system in Eq. (\ref{eq:ODEreduced}) as
\begin{equation}
\begin{array}{l}
x'_{HSC} = (2p_0^*-1)\eta_1 x_{HSC}, \\ \\
x'_{MPP} = 2(1-p_0^*)\eta_1 x_{HSC} -\eta_2^* x_{MPP},\\ 
\end{array}
\label{eq:ODEreduced}
\end{equation}
where 
\begin{equation}
\eta_2^*=\eta_2/(1+\gamma_2 x_{HSC}),
\label{new eta2}
\end{equation}
and $\eta_2=(1-2p_1)\tilde\eta_2$.

The non-linear feedback and feedforward loops in the above model enable the tight control of growth, the establishment of  equilibria that are robust to large changes in parameter values and rapid regeneration of equilibia after perturbations by external stimuli (e.g., \cite{Marciniak2009,Lander2009}), although the regeneration dynamics can exhibit oscillatory behavior. Finally, all the ODE model parameters can be grouped in the vector $\bm{\theta}\,=  \left(p_{0},\eta_{1},\eta_{2},\gamma_{1},\gamma_{2}\right)$.

\subsection*{Hierarchical Bayesian framework}

The goal of our statistical framework is to link the dynamic mathematical model with empirical data. 
We use a latent variables approach where the ODE model allows us to interpolate all the latent trajectories for the measurements that are missing before a mouse is harvested. 
According to the assumed data generation process described above, equations \eqref{eq:likelihoodt0} and  \eqref{eq:likelihoodt>0} define the likelihood function as the product of normal densities at time zero and at later times:
\begin{equation} 
\begin{array}{rl}
    p(\mathbf{y} \mid \bm{\Theta}) =
     &  \mathlarger{\prod}_{i=1}^k 
        N\left(\mathbf{y}_i  \mid \log\,\bm{\mu},
            \sigma_{t}^{2}+\sigma_{b}^{2}\right)
      \\
     &  \times  \mathlarger{\prod}_{j=k+1}^M 
        N\left(\mathbf{y}_j  \mid \log\,\mathbf{x}( \mathbf{u}_j, \bm{\theta},t_j),
            \sigma_{t}^{2}\right),
\end{array}
\label{eq:likelihood}
\end{equation}

where the vector $\bm{\Theta}=\left(\bm{\theta}, \mathbf{u}_{k+1:M},\sigma_{b}^{2},\sigma_{t}^{2},\bm{\mu}\right)^{t}$ includes all the parameters and latent variables of the hierarchical model, and $\mathbf{u}_{k+1:M}$ is a vector of latent initial cell counts of mice that are sacrificed after $t_0$.
Using a Bayesian approach, we provide measures of uncertainty to all model parameters  by calculating the posterior distribution of all the parameters:
\begin{equation}
  p(\bm{\Theta} \mid \mathbf{y}) \propto 
  p\left(\mathbf{y} \mid \bm{\Theta} \right) 
  p\left(\bm{\Theta}\right). 
\label{eq:posterior}
\end{equation}

We assume \textit{a priori} independence among the model parameters, resulting in the following prior distribution decomposition:

\begin{equation}
   p(\bm{\Theta}) 
   = p(\bm{\theta})
     \cdot p(\mathbf{u})
     \cdot p(\bm{\mu})
     \cdot p(\sigma_{b}^{2}) 
     \cdot p(\sigma_{t}^{2}).
\label{eq:priorsDecomp}
\end{equation}

Each of the parameters  $\theta_i\in\bm{\theta}$ has a log-normal prior:  $\log \theta_i \sim N(m_{\theta_i}, \sigma^2_{\theta_i})$, with the exception of 
 $p_0$ that requires a logit transformation:  $\textrm{logit}(\frac{p_0-\frac{1}{2}}{2}) \sim N(m_{p_0},\sigma^2_{p_0})$, 
since it is constrained between 0.5 and 1.
Here, $p(\mathbf{u})$ represents the mice initial conditions according to equation (\ref{eq:likelihoodlatentvars}), and $p(\bm{\mu})$ represents the prior on the mean of the initial conditions. 
Finally, $p(\sigma_{b}^{2})$ and $p(\sigma_{t}^{2})$ represent the prior distributions on the biological and technical noises correspondingly.
We describe how we approximate the posterior distribution in Eq. \eqref{eq:posterior} via Markov chain Monte Carlo (MCMC) in the Computational Implementation section below.

\paragraph{Experimental design}

In the laboratory, there are multiple variables that can be modified when an experiment is performed. 
Each combination of these variables defines an experimental design, which belongs to the set of all possible designs, $D$. 
The optimal experimental design goal is to discriminate between all the possible designs  by using a suitable utility metric $U(\mathbf{y},d)$ that quantifies the amount of information gain about the model parameters for a dataset $\mathbf{y}$ collected under the design $d\in D$.
The optimal design is determined by comparing the expected utility, where the expectation is taken over all datasets, $\mathbf{y}$, and model parameters, $\bm{\Theta}$, for the design $d$:
\begin{equation}
u(d)
=   \mathbb{E}_{\bm{\Theta},\mathbf{y}}
    \left[
        U(\mathbf{y},d)
    \right]
=   \intop_{\mathbf{y}}
    \intop_{\bm{\Theta}}
    U(\mathbf{y},d)
    p(\mathbf{y}|\bm{\Theta},d)
    p(\bm{\Theta})
    d\bm{\Theta} d\mathbf{y}.
\label{eq:MeanUtility}
\end{equation}

For the utility metric $U(\mathbf{y},d)$, we use the Kullback-Leibler (KL) divergence function that quantifies the information gain as the difference between the prior and posterior distributions of the model parameters: 
\begin{equation}
U(\mathbf{y},d) 
=   \intop_{\bm{\Theta}}
    \log\left(
        \frac{p(\bm{\Theta}|\mathbf{y},d)}{p(\bm{\Theta})}
    \right)
    p(\bm{\Theta}|\mathbf{y},d)
    d\bm{\Theta}.
\label{eq:KLDiv}
\end{equation}
This utility is equivalent to the Shannon information gain \cite{ryan2016review}. 
We take a similar approach to calculate the marginal expected utility for each parameter $\theta_j \in \bm{\Theta}$ individually: 
\begin{equation}
U_j(\mathbf{y},d) 
=   \intop_{\theta_j}
    \log\left(
        \frac{p_j(\theta_{j}|\mathbf{y},d)}{p(\theta_{j})}
    \right)
    p(\theta_j|\mathbf{y},d)
    d\theta_j.
\label{eq:KLDivIndiv}
\end{equation}

\subsection*{Computational Implementation}

\begin{figure}[t]
    \centering
    \includegraphics[width=0.95\textwidth]{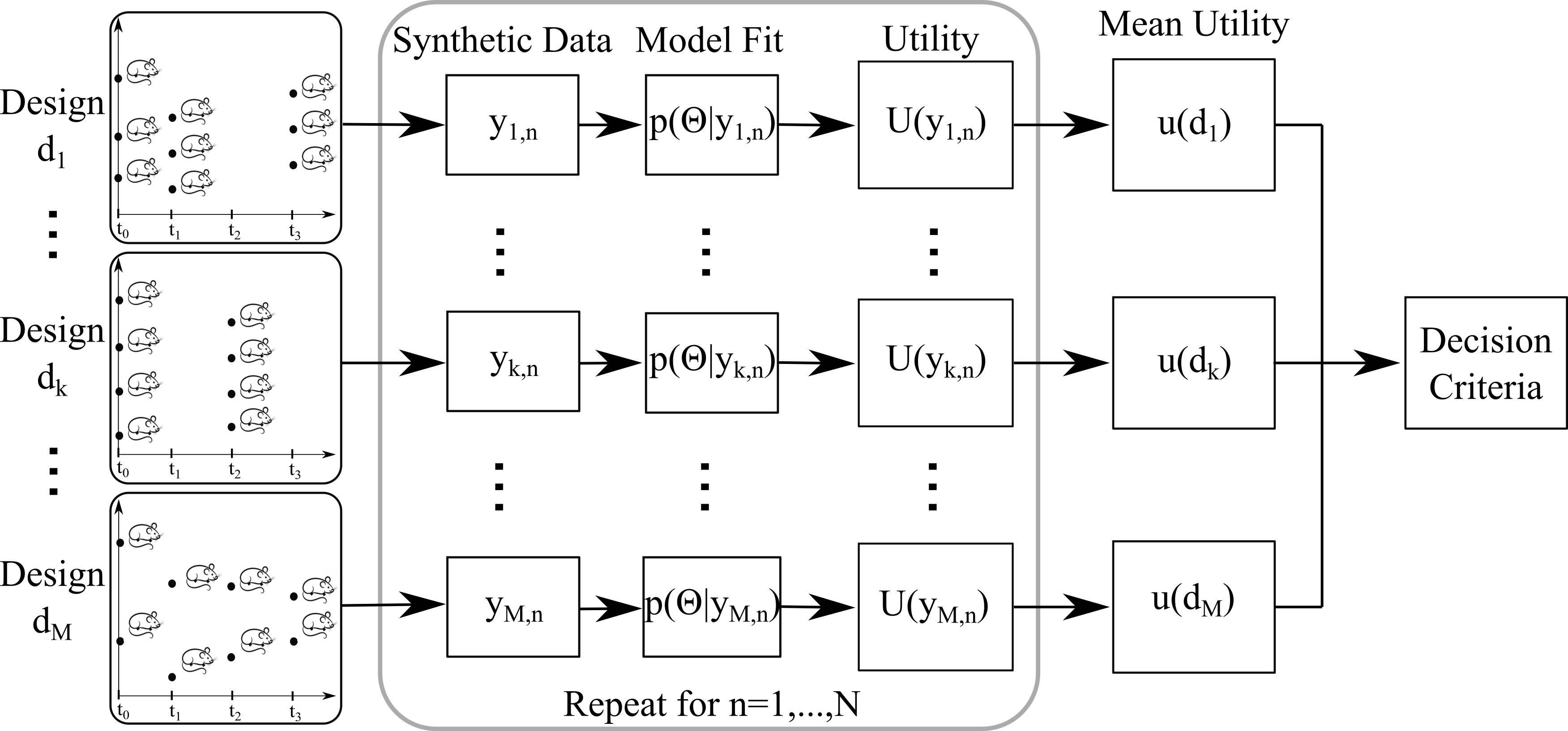}
    \caption{\textbf{Flow chart for computational implementation}. Given a design $d_k$, we sample a set of true parameter values  from the priors to obtain a synthetic data set generated from our mechanistic model using the latent variables approach. Using the synthetic data, we obtain the posterior distribution and we calculate the utility value associated to the $n$th synthetic data set. Once we have repeated this process $N$ times, we calculate the mean utility for design $d_k$. If we do this for $M$ designs, we are able to use this process to choose the design with the optimal mean utility based on a decision criterion that typically depends on several constraints. For example, these constraints can vary from minimizing the number of days or having a limit on the number of mice used.}  
  \label{fig:flow_chart}
  \vspace{-10pt}
\end{figure}

\paragraph{Estimating $p(\bm{\Theta} \mid \mathbf{y})$}
We approximate the posterior distribution in Eq. \eqref{eq:posterior} using an MCMC algorithm implemented in a statistical computing software platform called \texttt{Stan} \cite{stan2017, pystan2018}.
Stan provides the No U-Turn MCMC sampler, which allows the number of leapfrog steps during the warm-up phase of the MCMC simulation to be tuned.
This is especially useful in the case when parameters dependencies impose a complex geometry that makes the posterior sampling difficult. 
However, for our hierarchical model, we show in Section \ref{Sec:Results} that we can successfully obtain samples from the posterior distribution \eqref{eq:posterior}.

\paragraph{Estimating $U(\mathbf{y},d)$}
Each design $d\in D$ specifies the timing of the measurements and the number of mice replicates in a hematopoiesis experiment. 
When the design is fixed, we can simulate data by first sampling one set of true parameters from their corresponding prior distributions, and then sampling a synthetic dataset, $\mathbf{y}$, that contains the cellular population records for HSCs and MPPs according to the assumed data generating model.
At the initial time $t_0$, the initial conditions for the latent trajectories are determined by sampling from two independent normal distributions with means $\log(\mu_{HSC})$ and $\log(\mu_{MPP})$ and equal variances $\sigma_b^2 + \sigma_t^2$.
From the initial conditions, the ODE model \eqref{eq:ODEreduced} is forward-solved until the time when a trajectory is observed (mouse is sacrificed and the cell counts are observed). 
Then a random bivariate vector is generated from the normal distribution with the mean equal to the ODE solution and variance-covariance matrix $\mathbf{I} \sigma_t^2$. 
This process is repeated for all the latent trajectories to obtain the dataset $\mathbf{y}$.

Given a generated dataset, we obtain MCMC samples from the posterior distribution. The posterior samples are used to compute the value of the KL utility function in Eq. \ref{eq:MeanUtility} by Monte Carlo integration using all the MCMC iterations $i=1,\dots,n$:
\begin{equation}
\begin{array}{rrl}
U(\mathbf{y},d)\approx
&\widehat{U}(\mathbf{y},d) =
&  \frac{1}{n}
    \mathlarger{\sum}_{i=1}^{n}
    \log\left(
        \frac{p(\bm{\Theta}^{(i)}|\mathbf{y},d)}{p(\bm{\Theta}^{(i)})}
    \right) \\
 &=& \frac{1}{n}
    \mathlarger{\sum}_{i=1}^{n}
    \log \left(
        \frac{
        p(\mathbf{y} | \bm{\Theta}^{(i)}, d) 
            p(\bm{\Theta}^{(i)})}{
        p(\mathbf{y}|d)}
        
        \right) 
      - \log p(\bm{\Theta}^{(i)})
    \\
&= & \frac{1}{n}
    \mathlarger{\sum}_{i=1}^{n}
    \log p(\mathbf{y} | \bm{\Theta}^{(i)}, d) 
    - \log p(\mathbf{y}|d),
    \\
\end{array}
\label{eq:UtilityHat}
\end{equation}
where $\bm{\Theta}^{(i)}$ is the $i$-th sample from the posterior distribution which is evaluated at the prior and posterior densities. 
Note that the KL ratio in Eq. \eqref{eq:UtilityHat} requires an estimate of the marginal likelihood $p(\mathbf{y}|d)$, which  we obtain by Bridge Sampling --- a thermodynamics integration method for calculating the marginal likelihood  based on the posterior distribution samples \cite{bridge2017tutorial}.
The marginal KL utility value for each parameter $\theta_j$ is also approximated via Monte Carlo:
\begin{equation}
U_j(\mathbf{y},d)\approx \widehat{U}_j(\mathbf{y},d) 
= \frac{1}{n}
    \mathlarger{\sum}_{i=1}^{n}
    \log\left(
        \frac{\widehat{p}(\theta^{(i)}_j|\mathbf{y},d)}{p(\theta^{(i)}_j)}
    \right), 
\label{eq:UtilityHatIndividual}
\end{equation}

where $\widehat{p}(\theta_j|\mathbf{y},d)$ is an approximation for the true marginal posterior density and it is estimated by a Gaussian kernel density estimator using the posterior samples \cite{SciPy,scott2015multivariate}.
This step is required since the analytical marginal posterior density is generally not available.
Note that the kernel density estimation can only be applied to the marginal utility calculation since this approach is known to provide poor approximations for multiple-dimensional problems as in the case of the joint utility calculation.

\paragraph{Estimating $u(d)$.}

The overall mean utility in Eq. \eqref{eq:MeanUtility} is approximated by averaging dataset-specific utilities over $N$ simulated data:
\begin{equation}
    u(d)  
    \approx \widehat{u}(d) 
    = \frac{1}{N}\sum_{k=1}^N \widehat{U}(\mathbf{y}_k,d),
    \label{eq:MeanUtilityHat}
\end{equation}
where the utility $\widehat{U}(\mathbf{y}_k,d)$ is approximated using equation \eqref{eq:UtilityHat}. For individual parameters, the mean utility value is computed similarly:
\begin{equation}
    u_j(d)
    \approx \widehat{u}_j(d) 
    = \frac{1}{N}\sum_{k=1}^N \widehat{U}_j(\mathbf{y}_k,d),
    \label{eq:MeanUtilityHatIndividual}
\end{equation}
where $\widehat{U}_j(\mathbf{y}_k,d)$ is estimated using equation \eqref{eq:UtilityHatIndividual}.

\paragraph{Finding the optimal design} We consider a finite grid of designs $d_1, d_2,\dots,d_M$ that are relevant for the hematopoiesis experiment.
The optimal experimental design is determined by computing the expected utilities (\ref{eq:MeanUtilityHat})  and (\ref{eq:MeanUtilityHatIndividual}) for all the designs and finding a design with the maximum utility.
We illustrate the optimal design process in Figure \ref{fig:flow_chart}.

\section{Results}
\label{Sec:Results}

\paragraph{Successful parameter identification} 

\begin{figure}[t]
    \centering
    \includegraphics[width=0.95\textwidth]{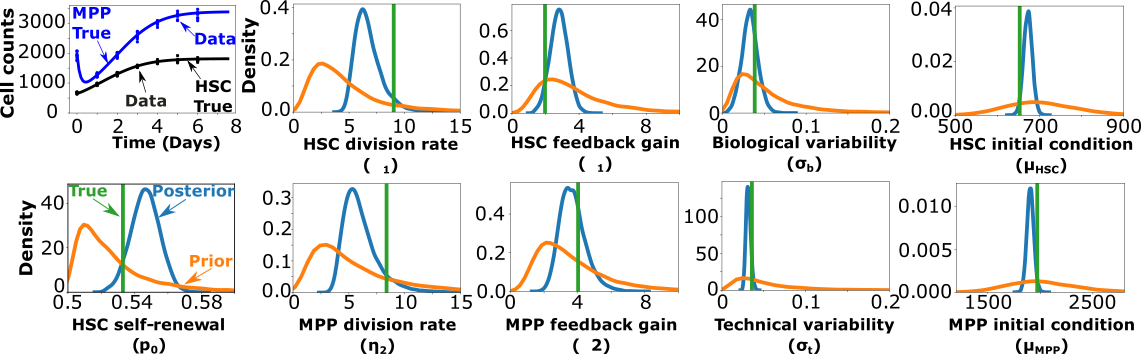}
     \caption{\textbf{Parameter identification for a synthetic data set.}
    Sampling from our prior distributions (see text), we generate a synthetic data set from our ODE solution (\textbf{Top Left}). 
    Solid curves correspond to ODE solutions for MPPs (blue) and HSCs (black). Symbols (Data) correspond to HSC and MPP cell numbers from the ODEs but with the addition of technical and biological noise following our latent variables framework (e.g., Eqs. \eqref{eq:likelihoodt0} and \eqref{eq:likelihoodt>0}). 
    The data corresponds to observations on  7 consecutive days with 7 mice replicates per day.  
    Using this synthetic data, in the remaining graphs we show the prior (orange) and posterior (blue) distributions for the ODE parameters and the latent HSC and MPP initial cell numbers together with the exact parameters and initial conditions used in the ODE model (green). 
    The ODE parameters, prior distributions and 95\% credible intervals are shown in the Supplemental Materials Table \ref{tab:PriorVsPosterior}.
    }
  \label{fig:SchematicSimStudy}
  \vspace{-8pt}
\end{figure}

Because our proposed mechanistic model is complex, we first need to ensure that the parameters in the model are identifiable. 
This is illustrated in Figure \ref{fig:SchematicSimStudy} using synthetic data. 
Here, we have generated a dataset, where the HSCs and MPPs are observed at 7 consecutive days with 7 replicates for each day (49 mice in total).
We have estimated the 5 ODE parameters, ($p_0$, $\eta_1$,$\eta_2$,$\gamma_1$, $\gamma_2$), the 2 initial conditions, ($\mu_{HSC}$ and $\mu_{MPP}$), the 2 error terms, ($\sigma_t$ and $\sigma_b$), and the 84 latent trajectories (93 parameters in total).
For this, we have used Stan \cite{stan2017,pystan2018} to obtain the posterior distribution of the model parameters. As we can see in Figure \ref{fig:SchematicSimStudy}, there are significant changes from prior to posterior distributions. 
Furthermore, we are able to recover the true parameter values within the 95\% posterior probability intervals.

To show that parameter identification is achieved under this design (7 days x 7 replicates), we have generated 60 synthetic datasets and obtained the posterior distribution for the model parameters. 
Using these results, we examine the coverage of the model parameters by calculating the percentage of the times the true parameter values are included in the 95\% credible intervals (equation \ref{eq:Coverage}). Additionally, we use the width of these intervals as a metric for the model precision (equation \ref{eq:MeanRelativeWidth}). 
We determine the relative bias for the MCMC simulations by using the posterior medians as point estimates and calculate the normalized residuals using the true parameter values (equation \ref{eq:MeanRelativeBias}). 
The estimated mean relative bias, mean relative width, and coverage of the 95\% credible intervals for all the parameters in the 7 days x 7 replicates design are shown in Table \ref{tab:coverage7x7}.
As we can see, most model parameters have good coverage probabilities. Also, except for $\gamma_2$ and $\sigma_b$, the mean relative bias tends to be small.

\begin{table}[t]
\centering
\begin{tabular}{cccccccccc}
\toprule 
7x7 Design  & $p_{0}$ & $\eta_{1}$ & $\eta_{2}$ & $\gamma_{1}$ & $\gamma_{2}$ & $\sigma_{t}$ & $\sigma_{b}$ & $\mu_{1}$ & $\mu_{2}$\tabularnewline
\midrule
Mean Relative Bias & {\small{}0.00} & {\small{}0.10} & {\small{}0.05} & {\small{}-0.08} & {\small{}-0.47} & {\small{}0.07} & {\small{}-0.31} & {\small{}-0.03} & {\small{}0.05}\tabularnewline
Mean Relative Width & {\small{}0.04} & {\small{}0.60} & {\small{}0.72} & {\small{}1.34} & {\small{}2.52} & {\small{}0.36} & {\small{}1.98} & {\small{}0.10} & {\small{}0.11}\tabularnewline
Coverage & {\small{}0.97} & {\small{}0.94} & {\small{}0.94} & {\small{}1.00} & {\small{}0.89} & {\small{}1.00} & {\small{}1.00} & {\small{}0.94} & {\small{}0.86}\tabularnewline
\bottomrule
\end{tabular}
\caption{\label{tab:coverage7x7}  Metrics for 60 simulations using 7 mice replicates during 7 days. The relative bias, relative width and coverage are calculated according to equations  \eqref{eq:MeanRelativeBias}, \eqref{eq:MeanRelativeWidth} and \eqref{eq:Coverage} shown on the Supplemental Materials.}
\end{table}

Tables \ref{tab:3ReplicatesCoverages} and \ref{tab:7ReplicatesCoverages} in the Supplemental Materials provide more simulation results based on other designs illustrated in Figure \ref{fig:flow_chart}.
In general, our results show that the true model parameters are included within the 95\% credible intervals. 
We also observe that designs with more data points allow for narrower posteriors, as quantified by the mean relative width, but higher bias is observed in some parameters. 

Next, we focus on quantifying information gain using the Bayesian utility theory described in the previous section.
In particular, we are interested in finding the optimal experimental setup that provides the highest information gain for inferring the model parameters.

\paragraph{Low dose radiation targets HSCs for cell death}
As a proof of concept, we use our hierarchical model to fit and obtain posterior distributions from preliminary data from a bone marrow perturbation experiment targeting stem cells.

In this experiment, low dose radiation (50 cGy) was applied to a number of mice following \cite{stewart1998lymphohematopoietic} where it was claimed that low dose radiation decreases HSC numbers. 
This is consistent with what we observed as well, see Fig. \ref{fig:pvp_HSC_IC} in the Supplemental Materials. To observe the system dynamics in response to this perturbation, the HSC and MPP cell numbers were obtained shortly after irradiation and at two other time points (days 0, 2 and 6). 
At each time point, some of the mice were sacrificed, their  bone marrows were extracted and sorted, and the cell numbers were determined by flow cytometry (see Experimental Methods in the Supplemental Materials). 
While there were 13 mice in total, seven  mice were sacrificed at day 0,  four were sacrificed on day 2 and two were sacrificed on day 6. 
Note the large variability in the data, particularly in the MPP numbers on Fig. \ref{fig:prasanthi_data}. 
The data suggests that the HSCs and MPPs return to equilibrium within one week. 

\begin{figure}[t]
    \centering
    \includegraphics[width=0.95\textwidth]{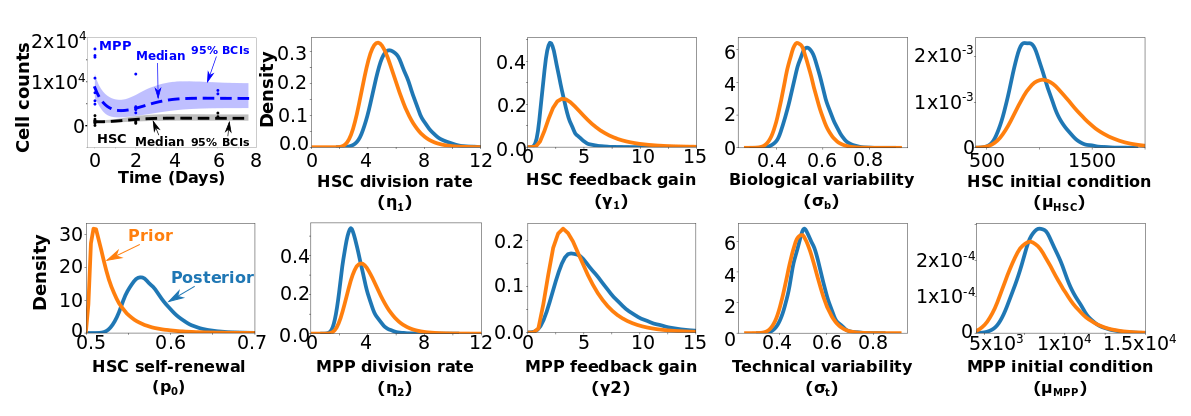}
     \caption{\textbf{Parameter estimation for a real data set.}
     \textbf{Top Left.} Preliminary data (symbols) obtained from mice that were irradiated with 50 cGy and fits of our hierarchical model to this dataset where we show the median (dashed) of the posterior distribution for the ODE solutions and the 95\% Bayesian credible intervals (bands).
    \textbf{Remaining graphs.} Plots of the prior and posterior distributions for the fitted ODE parameters and the mean of the latent HSC and MPP initial conditions. 
    Additional details on the priors and posterior distributions can be found on Supplemental Materials Table \ref{tab:FitDataPriorVsPosterior}.
    }
  \label{fig:fit_real_data}
  \vspace{-12pt}
\end{figure}

Fig. \ref{fig:fit_real_data}(top left) shows the fits of the mechanistic model to the data for HSCs (black) and MPPs (blue). 
In particular, the medians (dashed) of the posterior distribution of the ODE solutions are shown together with the 95$\%$ Bayesian credible regions \cite{Sivia2006} (shaded). 
The ODE fit seems reasonable based on the small amount of experimental data. 
However, only the HSCs initial conditions comparing  experiment vs. control data (Fig. \ref{fig:prasanthi_data}) showed a significant shift between the prior and posterior distributions (Fig. \ref{fig:pvp_HSC_IC}). 
All the other parameters did not show a substantial shift in the posterior distributions compared to the priors (Fig. \ref{fig:fit_real_data}, Table \ref{tab:FitDataPriorVsPosterior}, Fig. \ref{fig:pvp_gammas_sigmas}), suggesting there is little information gained from this data.
Moreover, the Bayes Factor (e.g., \cite{Sivia2006}) between a model with feedback (Eq. \eqref{eq:ODEreduced}) compared to the same model without feedback regulation (Eq. \eqref{eq:ODEreduced} with $\gamma_1$ and $\gamma_2$ identically zero) is equal to $BF=1.18$, which indicates that there is no strong evidence in favor of the model with feedback according to this limited dataset.

\begin{figure}[ht]
    \centering
    \includegraphics[width=0.95\textwidth]{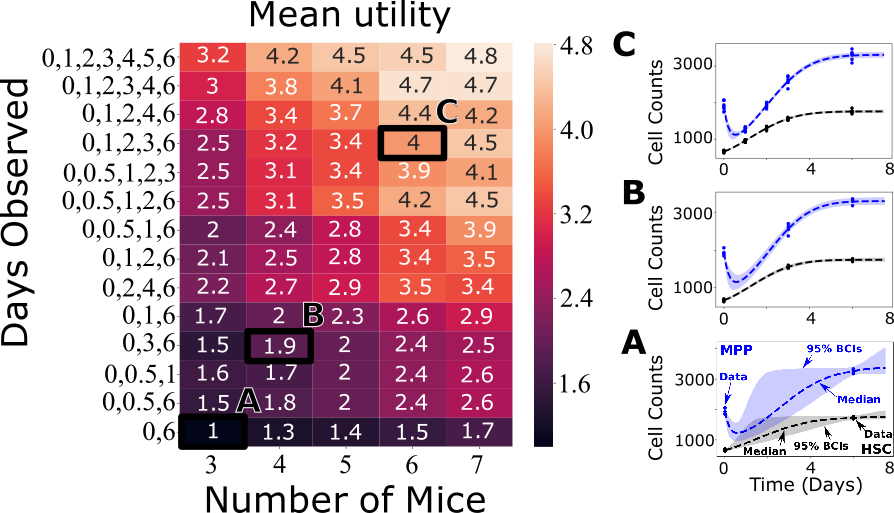}
    \caption{\textbf{Mean utility values for sampled designs.} \textbf{Left Column.} The mean utility heat map across all model parameters following the simulation process presented in Fig. \ref{fig:flow_chart}.
    Utility values correspond to a fold change to the minimum utility value, which corresponds to 3 mice observed at days 0 and 6. 
    The higher utility fold changes correspond to a greater information gain. 
    The boxes labeled \textbf{A}, \textbf{B} and \textbf{C} correspond to 3 designs with increasing mean utility value.
    \textbf{Right Column.} The medians and 95\% Bayesian Credible Intervals (bands) of the
    posterior distributions for the ODE solutions using the same synthetic dataset for the labeled designs.
    The parameter values additional details on the prior and posterior distributions
     are available in the Supplemental Materials Table \ref{tab:PriorVsPosterior}.
     }
  \label{fig:MeanUtilitiesHeatMap}
  \vspace{-8pt}
\end{figure}

\paragraph{Utility Grid Search}
Since we know that the model parameters are identifiable using a sufficient amount of synthetic data, the results in the previous section highlight the importance of finding a proper experimental design to maximize information gain in the radiation experiment. To this end, we explore different experimental setups by varying the number of mice collected per day and the timing of the measurements. 
That is, each design represents some number of mice observed at some sampling frequency.
We consider a finite number of experimental designs (70) to include a varying amount of mice over different observation days.
Our design space is defined as the following.
The first observation day starts at day 0 right after radiation, and more times are added until day 6 since it was shown in our preliminary perturbation experiment that the system returns to equilibrium in less than a week.
We also assume that the number of mice observed per day could be 3, 4, 5, 6, or 7. 

Exploring this finite experimental design space requires calculating the expected utilities for which we use 60 different synthetic datasets per design (\ref{eq:MeanUtilityHat}, \ref{eq:MeanUtilityHatIndividual}).
As shown in Figure \ref{fig:MeanUtilitiesHeatMap}, the expected utilities show an increasing trend when the number of replicates and the frequency of sampling increase. 
The values in this figure correspond to a fold change with respect to the baseline design, 3 mice observed at day 0 and 6, with minimum utility. 
To better understand the scale, we choose several designs (boxed), compare their mean utilities, and plot the corresponding ODE credible regions.
The plots provide a visual explanation on how higher information gain, as quantified by the mean utility values, correspond to designs with higher number of data points, which in turn provide narrower credible regions (Figure \ref{fig:MeanUtilitiesHeatMap}).
Note that higher mean utillity can be interpreted as lower uncertainty regarding the system dynamics. 

\paragraph{Parameter Utility}
The expected utility provides an overall metric for information gain by averaging over all model parameters.
Alternatively, to quantify the amount of information provided by the observed data with respect to a specific model parameter, we can use the individual parameter utilities. This way, we can better understand how data affects identification of certain model parameters.

\begin{figure}
    \centering
    \includegraphics[width=0.95\textwidth]{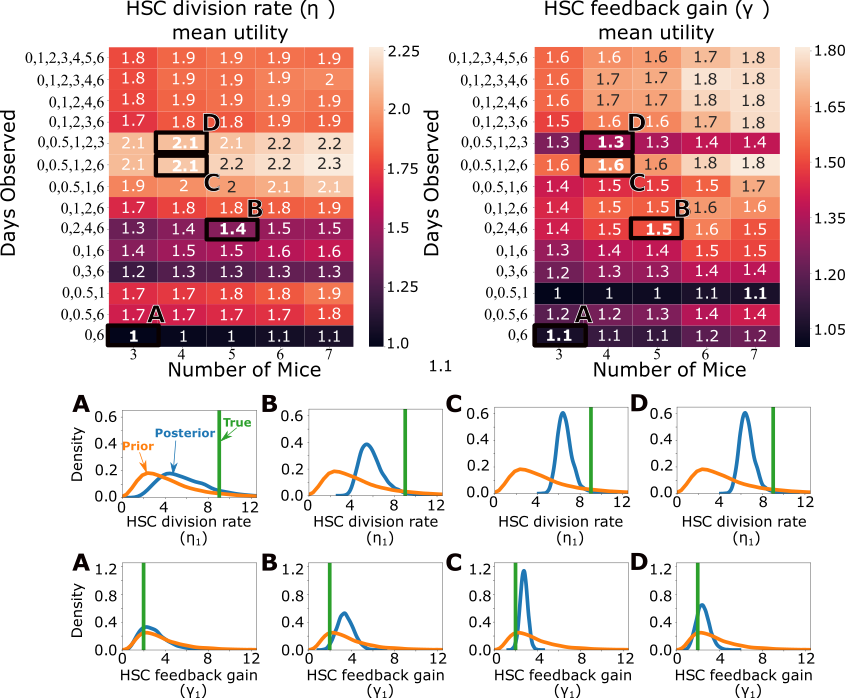}
    \caption{\textbf{Individual utilities for the HSC division rate and feedback gain.} 
    \textbf{Top Row}. Two mean utility heat maps for the HSC division rate $\eta_1$ and self-renewal feedback gain $\gamma_1$. The mean utility values are relative to the design with the lowest mean utility value for in each heat map.
    Four designs (labelled \textbf{A}-\textbf{D}) are selected with increasing mean utility values and different the sampling frequencies.
    \textbf{Bottom Row.} Plots of the prior and posterior distributions for $\eta_1$ and $\gamma_1$ for the selected experimental designs, as labelled. The true values are shown in green.
    }
  \label{fig:Gam1Eta1Utilities}
  \vspace{-8pt}
\end{figure}

Additionally, obtaining individual parameter utilities allows us to find a specific design that is more informative for a parameter of interest. 
For example, if we focus on SC division ($\eta_1$), the estimated parameter utilities suggest that more observations should be allocated to earlier days. 
Looking at the top four design rows for $\eta_1$ in Figure \ref{fig:Gam1Eta1Utilities}, the marginal mean utility value does not change even though we are adding more mice and more observation days. 
This shows that for $\eta_1$, the main contribution to information gain is coming from days 0 to 3. 
Similarly, if we focus on the HSC self-renewal probability feedback parameter, $\gamma_1$, we observe that adding data points at later times will lead to higher utilities. 
For the same parameter, we observe a lower marginal mean utility for designs where day 6 is not included. 
We also note that only a small amount of data is needed to identify the initial conditions, $\mu_1$ and $\mu_2$. 
For the full set of parameter utilities, please refer to Figure \ref{fig:heat_map_parameter_utils_all} in the Supplemental Materials.

As an example on how to use the parameter utilities for decision making, we can assume we have been given a finite budget of 20 mice for our perturbation experiment. Looking at the boxed B, C, and D marginal mean utilities in Figure \ref{fig:Gam1Eta1Utilities}, we can compare how the allocation of 20 mice could affect information gain for $\eta_1$ and $\gamma_1$. As mentioned previously, designs with more data points near the initial condition provide higher utilities for $\eta_1$ but lower utilities on $\gamma_1$. Design C provides a reasonable compromise between these two alternatives. For the full set of plots comparing prior and posterior distributions, please refer to Figure \ref{fig:pvp_param_util_all} in the Supplemental Materials. In general, we have found that given a fixed number of mice, designs with fewer mice at more time points tend to provide better results. Further, as we obtain more data, we see to reach a point of saturation. see Fig. \ref{fig:fixed_mice_comparison} in the Supplemental Materials for more design comparisons.

\section{Discussion \label{Sec:Discussion}}

We have presented a new method to find the optimal experimental design for inferring parameters in complex, mechanistic mathematical models where the experimental data is temporal but not longitudinal.That is, each data point in time corresponds to a different experimental subject. Our method incorporates Bayesian utility theory in a hierarchical latent variables framework where the mechanistic model is used to predict the unobserved temporal dynamics. We quantified the amount of information gained from a specific experimental design as the difference between the prior and posterior distributions of model parameters
using a utility metric based on the Kullback-Leibler divergence function \cite{kullback1997information,ryan2016review}. Further, by calculating marginal utilities, our proposed framework also allows us to identify the information gained for each parameter from a specific design. By searching over the space of possible experimental designs, the best design is the one that maximizes the expected information gain subject to specific constraints, e.g., the fixing the total number of experimental subjects.

We applied this framework to infer parameters in mathematical models of hematopoiesis in mice that incorporate feedback and feedforward regulation of self-renewal and division rates of hematopoietic stem cells (HSC) and multipotent progenitor (MPP) cells using experimental data on cell counts collected in mice. To obtain a cell count, each mouse had to be sacrificed and the bone marrow extracted and analyzed. In the experiments, mice were subjected to low doses of radiation, which decreased the number of HSCs in the bone marrow. After about seven days, the HSC and MPP cell counts returned to normal. 

We considered a finite grid of possible designs relevant for the radiation experiments where the control parameters were the number of mice observed at each time point and the time points at which the cell counts were obtained. Generally, we found that for designs with a fixed number of subjects, having more observation times with fewer mice replicates leads to higher information gain measured by the expected utility values. Additionally, we found that the amount of information gain from a specific design can vary significantly across parameters. For example, division rates are better informed by designs that include more measurements at early times, right after the radiation was applied, but the feedback parameters are better identified by designs that have more measurements at later times closer to the equilibrium. 

Our method relies on Monte Carlo averaging, MCMC simulation and Bridge sampling which can require a significant amount of time to implement for complex models. Faster sampling methods \cite{shahbaba2019deep, efendiev2005efficient, drovandi2018improving, beskos2017geometric}, efficient numerical solvers for differential equations \cite{calderhead2009accelerating, chkrebtii2016bayesian}, and a more efficient exploration of the design space \cite{ryan2014towards} can provide a significant reduction on computation time. 

From the modeling perspective, our framework can be extended to stochastic formulations of the hematopoietic system with explicit feedback terms that generalize well-known branching process models \cite{xu2019statistical}.  Incorporating such an approach would enable us to model the cellular stochasticities and the heterogeneity of the system components directly. However, fitting stochastic models tends to be difficult since the likelihood function is generally not available and approximate Bayesian methods have to be used. Future research directions could involve developing better implementations of stochastic process models for this problem. Our method could also be extended to incorporate data on more differentiated cell types in order to provide insight on the response dynamics for more realistic  models of the hematopoietic system with additional feedback mechanisms. 
To this end, we can include new perturbations, such as the depletion of  specific differentiated cells, into our approach to experimental design.

Additional experiments based on repeated measurements of covariates and cell counts can provide a better understanding of marginal and population-level responses of hematopoiesis regulation by allowing the model parameters change across individuals \cite{davidian2017nonlinear}.

Furthermore, our approach can be extended to incorporate richer sources of data such as serially sampled barcoded single cells, which allows for lineage tracking and fate determination. 

The hierarchical framework presented in this paper can significantly improve mathematical modeling of hematopoiesis by determining the required experiments to validate theoretical predictions and to obtain measures of uncertainty of the model components. 
In particular, our approach can be used for testing more complex feedback regulation and control mechanisms.
Finally, our work can motivate new collaborations between biologists, data scientists and mathematicians to develop a unified framework for hypothesis generation, modeling, and experimental validation \cite{navlakha2011algorithms}.

\section{Acknowledgements \label{Sec:Acknowledgements}}
 The authors acknowledge support from NSF grants DMS-1936833 (LML, AI, RVE, AL, BS, JSL, VM) and DMS-1714973 (JSL). In addition, AL, BS, JSL, VM acknowledge support from DMS-1763272 and the Simons Foundation (594598QN) for a NSF-Simons Center for Multiscale Cell Fate Research. RVE, AL, BS, JSL, VM also thank the National Institutes of Health for partial support through grants 1U54CA217378-01A1 for a National Center in Cancer Systems Biology at UC Irvine and P30CA062203 for the Chao Family Comprehensive Cancer Center at UC Irvine. Finally, LML acknowledges the support from the UC-MEXUS CONACYT and Fulbright-Garcia Robles doctoral fellowships.

\newpage
\bibliographystyle{unsrt}
\bibliography{references}

\begin{thebibliography}{10}

\bibitem{rieger2012hematopoiesis}
Rieger MA and Schroeder T.
\newblock Hematopoiesis.
\newblock {\em Cold Spring Harbor Perspectives in Biology}, 4:12, 2012.

\bibitem{SysBioBlood}
S.J. Corey, M.~Kimmel, and J.N. Leonard, editors.
\newblock {\em A Systems Biology Approach to Blood}, volume 844 of {\em
  Advances in Experimental Medicine and Biology}.
\newblock Springer, New York, USA, 2014.

\bibitem{Hofer2016}
T.~H\"ofer, M.~Barile, and M.~Flossdorf.
\newblock Stem-cell dynamics and lineage topology from in vivo fate mapping in
  the hematopoietic system.
\newblock {\em Curr. Opinion Biotechn.}, 39:150--156, 2016.

\bibitem{Pujo-Menjouet2016}
L.~Pujo-Menjouet.
\newblock Blood cell dynamics: Half of a century of modelling.
\newblock {\em Math. Model. Nat. Phenom.}, 11:92--115, 2016.

\bibitem{MacLean2017}
A.L. MacLean, C.~Lo~Celso, and M.P. Stumpf.
\newblock Concise review: Stem cell population biology: Insights from
  hematopoiesis.
\newblock {\em Stem Cells}, 35:80--88, 2017.

\bibitem{Fornari2018}
C.~Fornari, L.O. O'Connor, J.W.T. Yates, S.Y.A. Cheung, D.I. Jodrell, J.T.
  Mettetal, and T.A. Collins.
\newblock Understanding hematological toxicities using mathematical modeling.
\newblock {\em Clin. Pharmacol. Ther.}, 104:644--654, 2018.

\bibitem{michor2005dynamics}
Franziska Michor, Timothy~P Hughes, Yoh Iwasa, Susan Branford, Neil~P Shah,
  Charles~L Sawyers, and Martin~A Nowak.
\newblock Dynamics of chronic myeloid leukaemia.
\newblock {\em Nature}, 435(7046):1267, 2005.

\bibitem{Marciniak2009}
A~Marciniak-Czochra, T~Stiehl, AD~Ho, W~J{\"a}ger, and W~Wagner.
\newblock Modeling of asymmetric cell division in hematopoietic stem
  cells-regulation of self-renewal is essential for efficient repopulation.
\newblock {\em Stem Cells and Development}, 18(3):377--386, 2009.

\bibitem{busch2015fundamental}
Katrin Busch, Kay Klapproth, Melania Barile, Michael Flossdorf, Tim
  Holland-Letz, Susan~M Schlenner, Michael Reth, Thomas H{\"o}fer, and
  Hans-Reimer Rodewald.
\newblock Fundamental properties of unperturbed haematopoiesis from stem cells
  in vivo.
\newblock {\em Nature}, 518(7540):542, 2015.

\bibitem{Craig2016}
M.~Craig, A.R. Humphries, and M.C. Mackey.
\newblock A mathematical model of granulopoiesis incorporating the negative
  feedback dynamics and kinetics of g-csf/neutrophil binding and
  internalization.
\newblock {\em Bull. Math. Biol.}, 78:2304--2357, 2016.

\bibitem{Glauche2018}
I.~Glauche, M.~Kuhn, C.~Baldow, P.~Schulze, T.~Rothe, H.~Liebscher, A.~Roy,
  X.~Wang, and I.~Roeder.
\newblock Quantitative prediction of longterm molecular response in tkitreated
  cml ? lessons from an imatinib versus dasatinib comparison.
\newblock {\em Sci. Rep.}, 8:12330, 2018.

\bibitem{Mahadik2019computational}
Bhushan Mahadik, Bruce Hannon, and Brendan A.~C. Harley.
\newblock A computational model of feedback-mediated hematopoietic stem cell
  differentiation in vitro.
\newblock {\em PLOS ONE}, 14(3):1--21, 03 2019.

\bibitem{manesso2013dynamical}
Erica Manesso, Jos{\'e} Teles, David Bryder, and Carsten Peterson.
\newblock Dynamical modelling of haematopoiesis: an integrated view over the
  system in homeostasis and under perturbation.
\newblock {\em Journal of the Royal Society Interface}, 10(80):20120817, 2013.

\bibitem{roeder2002}
I.~Roeder and M.~Loeffler.
\newblock A novel dynamic model of hematopoietic stem cell organization based
  on the concept of within-tissue plasticity.
\newblock {\em Exp. Hematology}, 30:853--861, 2002.

\bibitem{Dingli2007}
D.~Dingli, A.~Traulsen, and J.M. Pacheco.
\newblock Stochastic dynamics of hematopoietic tumor stem cells.
\newblock {\em Cell Cycle}, 6(4):461--466, 2007.

\bibitem{Kimmel2014}
M.~Kimmel.
\newblock Stochasticity and determinism in models of hematopoiesis.
\newblock In S.J. Corey, M.~Kimmel, and J.N. Leonard, editors, {\em A Systems
  Biology Approach to Blood}, Advances in Experimental Medicine and Biology,
  pages 79--97. Springer, 2014.

\bibitem{Rozhik2016}
A.I. Rozhok, J.L. Salstrom, and J.~DeGregori.
\newblock Stochastic modeling reveals an evolutionary mechanism underlying
  elevated rates of childhood leukemia.
\newblock {\em Proc. Nat. Acad. Sci.}, 113(4):1050--1055, 2016.

\bibitem{Jakel2018}
J\"akel F., O.~Worm, S.~Lange, and R.~Mertrelsmann.
\newblock A stochastic model of myeloid cell lineages in hematopoiesis and
  pathway mutations in acute myeloid leukemia.
\newblock {\em PLoS One}, 13(10):e204393, 2018.

\bibitem{Xu2018}
J.~Xu, Y.~Wang, P.~Guttorp, and J.L. Abkowitz.
\newblock Visualizing hematopoiesis as a stochastic process.
\newblock {\em Blood Adv.}, 2(20):2637--2645, 2018.

\bibitem{krinner2013merging}
Axel Krinner, Ingo Roeder, Markus Loeffler, and Markus Scholz.
\newblock Merging concepts-coupling an agent-based model of hematopoietic stem
  cells with an ode model of granulopoiesis.
\newblock {\em BMC systems biology}, 7(1):117, 2013.

\bibitem{golinelli2006bayesian}
D~Golinelli, P~Guttorp, and JA~Abkowitz.
\newblock Bayesian inference in a hidden stochastic two-compartment model for
  feline hematopoiesis.
\newblock {\em Mathematical Medicine and Biology}, 23(3):153--172, 2006.

\bibitem{fong2009bayesian}
Youyi Fong, Peter Guttorp, and Janis Abkowitz.
\newblock Bayesian inference and model choice in a hidden stochastic
  two-compartment model of hematopoietic stem cell fate decisions.
\newblock {\em The annals of applied statistics}, 3(4):1696, 2009.

\bibitem{xu2019statistical}
Jason Xu, Samson Koelle, Peter Guttorp, Chuanfeng Wu, Cynthia~E Dunbar, Janis~L
  Abkowitz, and Vladimir~N Minin.
\newblock Statistical inference in partially observed stochastic compartmental
  models with application to cell lineage tracking of in vivo hematopoiesis.
\newblock {\em Annals of Applied Statistics}, 13:2091--2119, 2019.

\bibitem{oden2009toward}
J~Tinsley Oden, Andrea Hawkins, and Serge Prudhomme.
\newblock Toward predictive models of tumor growth: A general diffuse-interface
  continuum theory of mixtures for tumor growth and a bayesian approach to
  model validation and uncertainty quantification.
\newblock 2009.

\bibitem{farrell2017adaptive}
Kathryn Farrell-Maupin and JT~Oden.
\newblock Adaptive selection and validation of models of complex systems in the
  presence of uncertainty.
\newblock {\em Research in the Mathematical Sciences}, 4(1):14, 2017.

\bibitem{dehideniya2018optimal}
Mahasen~B Dehideniya, Christopher~C Drovandi, and James~M McGree.
\newblock Optimal bayesian design for discriminating between models with
  intractable likelihoods in epidemiology.
\newblock {\em Computational Statistics \& Data Analysis}, 124:277--297, 2018.

\bibitem{zhang2018optimal}
Jeff~F Zhang, Nikos~E Papanikolaou, Theodore Kypraios, and Christopher~C
  Drovandi.
\newblock Optimal experimental design for predator--prey functional response
  experiments.
\newblock {\em Journal of The Royal Society Interface}, 15(144):20180186, 2018.

\bibitem{han2004bayesian}
Cong Han and Kathryn Chaloner.
\newblock Bayesian experimental design for nonlinear mixed-effects models with
  application to hiv dynamics.
\newblock {\em Biometrics}, 60(1):25--33, 2004.

\bibitem{kullback1997information}
S.~Kullback and R.A. Leibler.
\newblock On information and sufficiency.
\newblock {\em Annals Math. Stat.}, 22:79--86, 1951.

\bibitem{ryan2016review}
Elizabeth~G Ryan, Christopher~C Drovandi, James~M McGree, and Anthony~N
  Pettitt.
\newblock A review of modern computational algorithms for bayesian optimal
  design.
\newblock {\em International Statistical Review}, 84(1):128--154, 2016.

\bibitem{chaloner1995bayesian}
Kathryn Chaloner and Isabella Verdinelli.
\newblock Bayesian experimental design: A review.
\newblock {\em Statistical Science}, pages 273--304, 1995.

\bibitem{lindley1956measure}
Dennis~V Lindley et~al.
\newblock On a measure of the information provided by an experiment.
\newblock {\em The Annals of Mathematical Statistics}, 27(4):986--1005, 1956.

\bibitem{cook2008optimal}
Alex~R Cook, Gavin~J Gibson, and Christopher~A Gilligan.
\newblock Optimal observation times in experimental epidemic processes.
\newblock {\em Biometrics}, 64(3):860--868, 2008.

\bibitem{huan2013simulation}
Xun Huan and Youssef~M Marzouk.
\newblock Simulation-based optimal bayesian experimental design for nonlinear
  systems.
\newblock {\em Journal of Computational Physics}, 232(1):288--317, 2013.

\bibitem{muller1995optimal}
Peter M{\"u}ller and Giovanni Parmigiani.
\newblock Optimal design via curve fitting of monte carlo experiments.
\newblock {\em Journal of the American Statistical Association},
  90(432):1322--1330, 1995.

\bibitem{wakefield1994expected}
Jon Wakefield.
\newblock An expected loss approach to the design of dosage regimens via
  sampling-based methods.
\newblock {\em Journal of the Royal Statistical Society: Series D (The
  Statistician)}, 43(1):13--29, 1994.

\bibitem{palmer1998bayesian}
J~Lynn Palmer and Peter M{\"u}ller.
\newblock Bayesian optimal design in population models for haematologic data.
\newblock {\em Statistics in Medicine}, 17(14):1613--1622, 1998.

\bibitem{drovandi2013sequential}
Christopher~C Drovandi, James~M McGree, and Anthony~N Pettitt.
\newblock Sequential monte carlo for bayesian sequentially designed experiments
  for discrete data.
\newblock {\em Computational Statistics \& Data Analysis}, 57(1):320--335,
  2013.

\bibitem{muller1999simulation}
P~M{\"u}ller.
\newblock Simulation-based optimal design.
\newblock {\em Bayesian Statistics}, 6, 1999.

\bibitem{biegler2011large}
Lorenz Biegler, George Biros, Omar Ghattas, Matthias Heinkenschloss, David
  Keyes, Bani Mallick, Luis Tenorio, Bart van Bloemen~Waanders, Karen Willcox,
  and Youssef Marzouk.
\newblock {\em Large-scale inverse problems and quantification of uncertainty},
  volume 712.
\newblock John Wiley \& Sons, 2011.

\bibitem{muller2006bayesian}
Peter M{\"u}ller, Don~A Berry, Andrew~P Grieve, and Michael Krams.
\newblock A bayesian decision-theoretic dose-finding trial.
\newblock {\em Decision analysis}, 3(4):197--207, 2006.

\bibitem{liepe2013maximizing}
Juliane Liepe, Sarah Filippi, Micha{\l} Komorowski, and Michael~PH Stumpf.
\newblock Maximizing the information content of experiments in systems biology.
\newblock {\em PLoS Comput Biol}, 9(1):e1002888, 2013.

\bibitem{silk2014model}
Daniel Silk, Paul~DW Kirk, Chris~P Barnes, Tina Toni, and Michael~PH Stumpf.
\newblock Model selection in systems biology depends on experimental design.
\newblock {\em PLoS Comput Biol}, 10(6):e1003650, 2014.

\bibitem{quesenberry1998}
Stewart FM, Zhong S, Wuu J, Hsieh C-C, Nilsson SK, and Quesenberry PJ.
\newblock Lymphohematopoietic engraftment in minimally myeloablated hosts.
\newblock {\em Blood}, 91:3681--3687, 1998.

\bibitem{AbdonThesis}
A.~Inigiuez.
\newblock Mathematical modeling of malignant myelopoiesis: Optimal experimental
  design and targeted therapy.
\newblock {\em Ph.D. Thesis, University of California, Irvine}, 2019.

\bibitem{Lander2009}
Arthur~D Lander, Kimberly~K Gokoffski, Frederic Y.~M Wan, Qing Nie, and Anne~L
  Calof.
\newblock Cell lineages and the logic of proliferative control.
\newblock {\em PLOS Biology}, 7(1):1--1, 01 2009.

\bibitem{Buzi2015}
Gentian Buzi, Arthur~D. Lander, and Mustafa Khammash.
\newblock Cell lineage branching as a strategy for proliferative control.
\newblock {\em BMC Biology}, 13(1):13, Feb 2015.

\bibitem{Dharampuriya2017}
P.R. Dharampuriya, G.~Scapin, C.~Wong, K.J. Wagner, J.L. Cillis, and
  S.~Dhavanit.
\newblock Tracking the origin, development, and differentiation of
  hematopoietic stem cells.
\newblock {\em Curr. Opinion Cell Biol.}, 49:108--115, 2017.

\bibitem{Brown2018}
G.~Brown, P.~Tsapogas, and R.~Ceredig.
\newblock The changing face of hematopoiesis: A spectrum of options is
  available to stem cells.
\newblock {\em Immunol. Cell Biol.}, 96:898--911, 2018.

\bibitem{olsson2016single}
Andre Olsson, Meenakshi Venkatasubramanian, Viren~K Chaudhri, Bruce~J Aronow,
  Nathan Salomonis, Harinder Singh, and H~Leighton Grimes.
\newblock Single-cell analysis of mixed-lineage states leading to a binary cell
  fate choice.
\newblock {\em Nature}, 537(7622):698, 2016.

\bibitem{Arai2004}
F.~Arai, A.~Hirao, M.~Ohmura, H.~Sato, S.~Matsuoka, K.~Takubo, K.~Ito, and
  J.-Y.~and Koh.
\newblock Tie2/angiopoietin-1 signaling regulates hematopoietic stem cell
  quiescence in the bone marrow niche.
\newblock {\em Cell}, 118:149--161, 2004.

\bibitem{Staversky2018}
R.J. Staversky, D.K. Byun, M.A. Georger, B.J. Zaffuto, A.~Goodman, M.W. Becker,
  L.M. Calvi, and B.J. Frisch.
\newblock The chemokine ccl3 regulates myeloid differentiation and
  hematopoietic stem cell numbers.
\newblock {\em Sci. Rep.}, 8:14691, 2018.

\bibitem{stan2017}
Bob Carpenter, Andrew Gelman, Matthew~D Hoffman, Daniel Lee, Ben Goodrich,
  Michael Betancourt, Marcus Brubaker, Jiqiang Guo, Peter Li, and Allen
  Riddell.
\newblock Stan: A probabilistic programming language.
\newblock {\em Journal of statistical software}, 76(1), 2017.

\bibitem{pystan2018}
{Stan Development Team}.
\newblock {PyStan: the Python interface to Stan. Version 2.17.1.0}, 2018.

\bibitem{bridge2017tutorial}
Quentin~F Gronau, Alexandra Sarafoglou, Dora Matzke, Alexander Ly, Udo Boehm,
  Maarten Marsman, David~S Leslie, Jonathan~J Forster, Eric-Jan Wagenmakers,
  and Helen Steingroever.
\newblock A tutorial on bridge sampling.
\newblock {\em Journal of mathematical psychology}, 81:80--97, 2017.

\bibitem{SciPy}
Eric Jones, Travis Oliphant, Pearu Peterson, et~al.
\newblock {SciPy}: Open source scientific tools for {Python}, 2001--.
\newblock [Online; accessed <today>].

\bibitem{scott2015multivariate}
David~W Scott.
\newblock {\em Multivariate density estimation: theory, practice, and
  visualization}.
\newblock John Wiley \& Sons, 2015.

\bibitem{stewart1998lymphohematopoietic}
FM~Stewart, S~Zhong, J~Wuu, C-c Hsieh, SK~Nilsson, and PJ~Quesenberry.
\newblock Lymphohematopoietic engraftment in minimally myeloablated hosts.
\newblock {\em Blood}, 91(10):3681--3687, 1998.

\bibitem{Sivia2006}
D.S. Silva and J.~Skilling.
\newblock {\em Data Analysis: A Bayesian tutorial}.
\newblock Oxford U. Press, 2006.

\bibitem{shahbaba2019deep}
Babak Shahbaba, Luis~Martinez Lomeli, Tian Chen, and Shiwei Lan.
\newblock Deep markov chain monte carlo, 2019.

\bibitem{efendiev2005efficient}
Yalchin Efendiev, Akhil Datta-Gupta, Victor Ginting, Xiang Ma, and Bani
  Mallick.
\newblock An efficient two-stage markov chain monte carlo method for dynamic
  data integration.
\newblock {\em Water Resources Research}, 41(12), 2005.

\bibitem{drovandi2018improving}
Christopher~C Drovandi, Minh-Ngoc Tran, et~al.
\newblock Improving the efficiency of fully bayesian optimal design of
  experiments using randomised quasi-monte carlo.
\newblock {\em Bayesian Analysis}, 13(1):139--162, 2018.

\bibitem{beskos2017geometric}
Alexandros Beskos, Mark Girolami, Shiwei Lan, Patrick~E Farrell, and Andrew~M
  Stuart.
\newblock Geometric mcmc for infinite-dimensional inverse problems.
\newblock {\em Journal of Computational Physics}, 335:327--351, 2017.

\bibitem{calderhead2009accelerating}
Ben Calderhead, Mark Girolami, and Neil~D Lawrence.
\newblock Accelerating bayesian inference over nonlinear differential equations
  with gaussian processes.
\newblock In {\em Advances in neural information processing systems}, pages
  217--224, 2009.

\bibitem{chkrebtii2016bayesian}
Oksana~A Chkrebtii, David~A Campbell, Ben Calderhead, Mark~A Girolami, et~al.
\newblock Bayesian solution uncertainty quantification for differential
  equations.
\newblock {\em Bayesian Analysis}, 11(4):1239--1267, 2016.

\bibitem{ryan2014towards}
Elizabeth~G Ryan, Christopher~C Drovandi, M~Helen Thompson, and Anthony~N
  Pettitt.
\newblock Towards bayesian experimental design for nonlinear models that
  require a large number of sampling times.
\newblock {\em Computational Statistics \& Data Analysis}, 70:45--60, 2014.

\bibitem{davidian2017nonlinear}
Marie Davidian.
\newblock {\em Nonlinear models for repeated measurement data}.
\newblock Routledge, 2017.

\bibitem{navlakha2011algorithms}
Saket Navlakha and Ziv Bar-Joseph.
\newblock Algorithms in nature: the convergence of systems biology and
  computational thinking.
\newblock {\em Molecular systems biology}, 7(1):546, 2011.

\end{thebibliography}

\newpage

\setcounter{page}{1}
\setcounter{table}{0}
\renewcommand{\thetable}{A-\arabic{table}}
\renewcommand{\thefigure}{A-\arabic{figure}}
\renewcommand{\thesection}{A-\arabic{section}}

\renewcommand{\theequation}{A-\arabic{equation}}
\setcounter{equation}{0}
\setcounter{section}{0}
\setcounter{figure}{0}

\newpage
\section*{Supplemental Materials\label{Sec:Supplemental}}

\subsection*{Experimental methods}

\paragraph{Mice} C57B/6J female mice (Jackson Laboratories), 6-12 weeks of age were used for irradiation and myeloid depletion experiments. All protocols in mouse was approved by Institutional Animal Use and Care Committee of University of California, Irvine.

\paragraph{Irradiation of mice} To achieve selective depletion of Hematopoietic Stem Cells (HSC), a 50cGy dose of irradiation from an x-ray source was applied. Control mice did not receive irradiation. 

\paragraph{Flow cytometry analysis of cell populations} Bone marrow (BM) cells from femur and tibia of control and dosed mice were isolated by flushing bones. RBC lysed cells (RBC lysis buffer, ebiosciences) were stained with CD34 antibody for one hour and subsequently incubated with a cocktail of biotinylated lineage markers CD3, Gr1, B220, Ter119, and c-kit, sca1, CD48 for 30 minutes. Streptavidin (SA) conjugated fluorchrome was utilized to detect biotynalated antibodies.  Following fix/ permeabilization and Dnase digestion, anti-BrdU antibody was used to check BrdU incorporation. Cells were acquired on FACS Arial II and analyzed with Flowjo v10 software

\paragraph{Antibodies} Monoclonal antibodies for flow cytometry were biotin mouse lineage panel (559971, BD biosciences), PE-CF594 Streptavidin (562318, BD biosciences), anti-mouse CD48 (561242, BD biosciences), anti-mouse CD34 eFluor450 (48-0341-82, ebiosciences), anti-mouse Sca1 PE (108108, Biolegend), anti-mouse c-Kit-APC (17-1171-82, ebiosciences), FITC BrdU flow kit (552598, BD biosciences)

\begin{figure}[ht]
    \centering
    \includegraphics[width=0.7\textwidth]{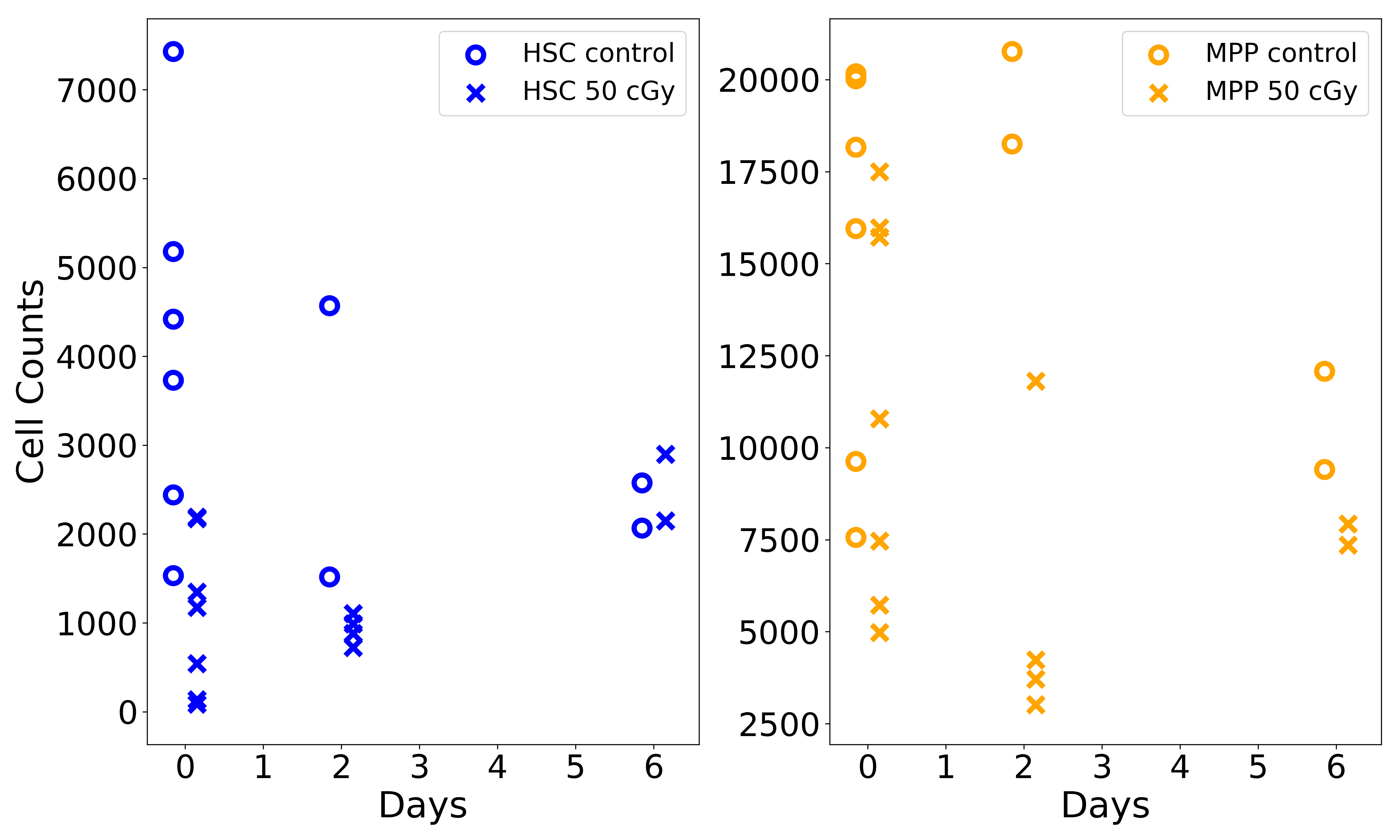}
    \caption{Perturbation experiment dataset. 
    HSCs and MPPs cellular counts are shown for three different time points. Dots and marks are use to denote control and perturbation (50 cGy radiation) experiments respectively. 
    Each point represents a single mouse data.}
    \label{fig:prasanthi_data}
\end{figure}

\begin{figure}[ht]
    \centering
    \includegraphics[width=0.5\textwidth]{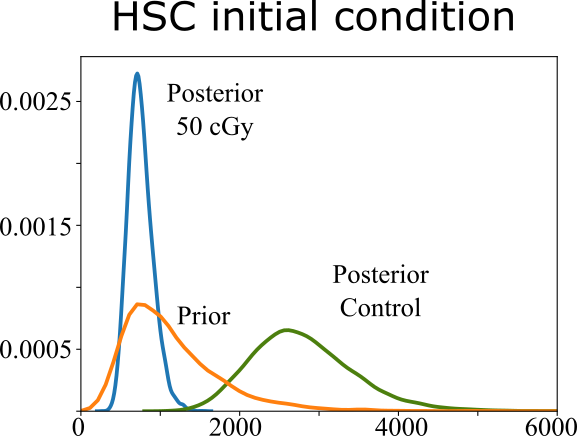}
    \caption{Prior and posterior distributions for the mean HSC control and perturbation initial conditions using the perturbation experiment dataset.}
    \label{fig:pvp_HSC_IC}
\end{figure}

\begin{figure}[ht]
    \centering
    \includegraphics[width=0.3\textwidth]{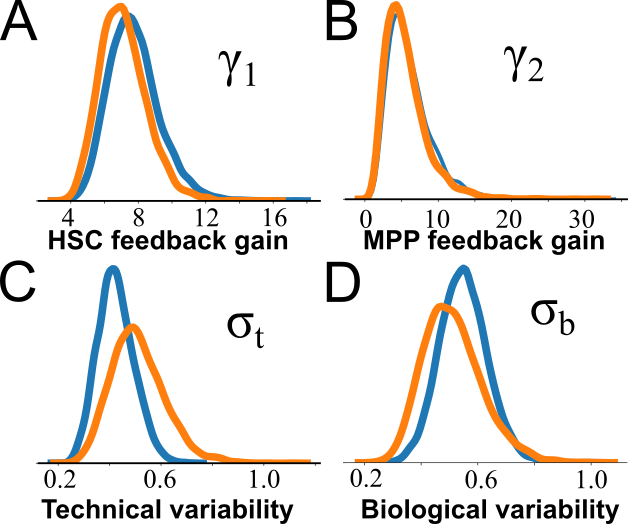}
    \caption{Prior and posterior distributions for the feedback gain ODE parameters (\textbf{A}, \textbf{B}) and the technical (\textbf{C}) and biological (\textbf{D}) variability parameters using the perturbation experiment dataset.}
    \label{fig:pvp_gammas_sigmas}
\end{figure}

\begin{table}[ht]
\centering
\begin{tabular}{ c|ccc|ccc|ccc}
 \toprule
 \multicolumn{1}{c}{} & & Prior & & &Posterior&  \\
 \midrule
 Parameter & 2.5\% & 50\% & 97.5\% & 2.5\% & 50\% & 97.5\% & \\ 
 \midrule
 $p_0$   & 0.50    & 0.52 &   0.60 & 0.54 & 0.57 & 0.64 \\
$\eta_1$ & 3.06 & 5 & 8.16 & 3.79 &  5.85 & 9.07  \\
 $\eta_2$ & 2.14 & 3.86 & 6.94 & 1.84 & 3.01 & 5.07   \\
 $\gamma_1$    & 1.50  & 4 & 10.7 & 1.06  & 2.29  &  4.88 \\
 $\gamma_2$ & 1.50  & 4 & 10.7 & 1.89  & 5.09  & 12.9   \\
 \midrule
 $\sigma_t$& 0.39 & 0.5 & 0.64 & 0.41  & 0.52 &  0.65  \\
 $\sigma_b$& 0.39 & 0.5 & 0.64  & 0.42  & 0.54 & 0.68  \\
 \midrule
 $\mu_1$ & 672 & 1097& 1791 & 634  & 915 & 1317   \\
 $\mu_2$ & 5473& 8103& 11988& 6280  & 8653  & 11810   \\
 \bottomrule

\end{tabular}
\caption{\label{tab:FitDataPriorVsPosterior} 
Prior vs posterior distributions for the model parameters using the perturbation experiment dataset.  
The 2.5\%, 50\% and 97.5\% percentiles are provided.}
\end{table}

\begin{table}[ht]
\centering
\begin{tabular}{ c|ccc|ccc|cccc  }
 \toprule
 \multicolumn{1}{c}{} & & Prior & & &Posterior& & True \\
 \midrule
 Parameter & 2.5\% & 50\% & 97.5\% & 2.5\% & 50\% & 97.5\% & \\ 
 \midrule
 $p_0$   & 0.50    & 0.52 &   0.60 & 0.53 & 0.55 & 0.54 & 0.53 \\
$\eta_1$ & 1.00 & 3.84 & 15.11 & 5.00 & 6.51 & 9.85 & 9.02 \\
 $\eta_2$ & 1.00 & 4.50 & 20.2 & 3.98 & 5.72& 10.1 & 8.37 \\
 $\gamma_1$    & 1.00  & 3.15 & 10 & 1.77 & 2.81 & 3.84 & 1.97\\
 $\gamma_2$ & 1.00  & 3.15 & 10 & 2.44 & 3.64 & 5.51  & 4.03 \\
 \midrule
 $\sigma_t$& 0.01 & 0.04 & 0.18 & 0.01 & 0.03&  0.05 & 0.04 \\
 $\sigma_b$& 0.01 & 0.04 & 0.18  &  0.03 &  0.03 & 0.04 & 0.04 \\
 \midrule
 $\mu_1$ & 549 & 700& 890& 652 & 672& 693 & 653 \\
 $\mu_2$ & 1462& 1996& 2743& 1839 & 1904 & 1970 & 1970 \\
 \bottomrule

\end{tabular}
\caption{\label{tab:PriorVsPosterior} Prior vs posterior distributions for the model parameters using a synthetic dataset with 7 replicates for 7 consecutive days. 
The 2.5\%, 50\% and 97.5\% percentiles are provided and the true parameter values are shown on the rightmost column.}
\end{table}

\paragraph{Relative bias, relative width and coverage}
In our simulation study, for each design $d$ we used several synthetic datasets and their corresponding MCMC simulation to infer back the true parameter values. For each of these simulations, we calculated the mean relative bias, mean relative width of the credible intervals and the coverage. 

First, we use the posterior median as a point estimate for each parameter $i$ in the hierarchical model for the $j$-th simulation run,

\begin{equation}
    \hat{\theta}_{i,j}^{(d)} = 
        median(
            p(\theta_{i,j}
                | y_j^{(d)}))
\end{equation}

where $y_j^{(d)}$ represents the $j$-th synthetic dataset for the design $d$.
The relative bias is calculated by normalizing the difference between the posterior median  and the true value, ie, for the i-th parameter

\begin{equation}
    \textrm{RelBias}^{(d)}_{i,j}  =
    \frac{
        \hat{\theta}_{i,j}^{(d)}
        - \theta_{i,j} 
        }{\theta_{i,j}}
        \label{eq:RelativeBias}
\end{equation}

where $\theta_{\cdot,j}$ represents the true parameter values for simulation $j$.
The mean relative bias is calculated for each design averaging over all the relative bias estimates for all the simulations of design $d$ as

\begin{equation}
    \textrm{Mean RelBias}_{i}^{(d)} = 
        \frac{1}{N} 
            \sum_{j=1}^{N} \textrm{RelBias}^{(d)}_{i,j}
    \label{eq:MeanRelativeBias}
\end{equation}

We define the coverage of a credible interval with a similar interpretation as a frequentist confidence interval is calculated as the proportion of the time that the interval contains the true value of interest. In this case we calculate the number of times the true parameter value $\theta_{i,j}$ is included in the corresponding 95\% credible interval $C_{i,j}$ for the design $d$. 

\begin{equation}
    \textrm{Coverage}^{(d)}_i = 
    \frac{1}{N} 
        \sum_{j=1}^{N} \mathbb{1}_{\theta_{i,j}\in C_{i,j}}
    \label{eq:Coverage}
\end{equation}

The relative width of the posterior distribution of a parameter is given by the magnitude of the 95\% credible interval, ie, the distance between the 97.5th and 2.5th percentiles as: 
\begin{equation}
    \textrm{RelWidth}^{(d)}_{i,j} =
    q_{0.975}(p(\theta_{i,j} | y_j^{(d)})) - q_{0.025}(p(\theta_{i,j} | y_j^{(d)})) 
    \label{eq:RelativeWidth}
\end{equation}

and the mean relative width is given by 
\begin{equation}
    \textrm{Mean RelWidth}^{(d)}_i = 
    \frac{1}{N}
        \sum_{j=1}^{N}
            \textrm{RelWidth}^{(d)}_{i,j}
    \label{eq:MeanRelativeWidth}
\end{equation}


\begin{table}[ht]
\resizebox{\columnwidth}{!}{%
    \addtolength{\leftskip} {-2cm}
    \addtolength{\rightskip}{-2cm}

\begin{tabular}{cccccccccccccccc}
\toprule 
\begin{turn}{90}
{\scriptsize{}Parameter}
\end{turn} & \begin{turn}{90}
{\scriptsize{}3 replicates}
\end{turn} & \begin{turn}{90}
{\scriptsize{}0,6}
\end{turn} & \begin{turn}{90}
{\scriptsize{}0,0.5,6}
\end{turn} & \begin{turn}{90}
{\scriptsize{}0,0.5,1}
\end{turn} & \begin{turn}{90}
{\scriptsize{}0,3,6}
\end{turn} & \begin{turn}{90}
{\scriptsize{}0,1,6}
\end{turn} & \begin{turn}{90}
{\scriptsize{}0,2,4,6}
\end{turn} & \begin{turn}{90}
{\scriptsize{}0,1,2,6}
\end{turn} & \begin{turn}{90}
{\scriptsize{}0,0.5,1,6}
\end{turn} & \begin{turn}{90}
{\scriptsize{}0,0.5,1,2,6}
\end{turn} & \begin{turn}{90}
{\scriptsize{}0,0.5,1,2,3}
\end{turn} & \begin{turn}{90}
{\scriptsize{}0,1,2,3,6}
\end{turn} & \begin{turn}{90}
{\scriptsize{}0,1,2,4,6}
\end{turn} & \begin{turn}{90}
{\scriptsize{}0,1,2,3,4,6}
\end{turn} & \begin{turn}{90}
{\scriptsize{}0,1,2,3,4,5,6}
\end{turn}\tabularnewline
\midrule 
\multirow{3}{*}{$p_{0}$} & {\tiny{}Mean RelBias} & {\scriptsize{}0.0} & {\scriptsize{}-0.0} & {\scriptsize{}-0.0} & {\scriptsize{}0.0} & {\scriptsize{}0.01} & {\scriptsize{}0.0} & {\scriptsize{}0.0} & {\scriptsize{}-0.0} & {\scriptsize{}-0.0} & {\scriptsize{}-0.01} & {\scriptsize{}0.0} & {\scriptsize{}0.0} & {\scriptsize{}-0.0} & {\scriptsize{}-0.0}\tabularnewline
 & {\tiny{}Coverage} & {\scriptsize{}0.91} & {\scriptsize{}0.97} & {\scriptsize{}1.0} & {\scriptsize{}0.96} & {\scriptsize{}0.97} & {\scriptsize{}0.93} & {\scriptsize{}0.93} & {\scriptsize{}1.0} & {\scriptsize{}0.93} & {\scriptsize{}0.89} & {\scriptsize{}0.92} & {\scriptsize{}0.91} & {\scriptsize{}0.92} & {\scriptsize{}0.93}\tabularnewline
 & {\tiny{}Mean RelWidth} & {\scriptsize{}0.05} & {\scriptsize{}0.04} & {\scriptsize{}0.05} & {\scriptsize{}0.04} & {\scriptsize{}0.04} & {\scriptsize{}0.03} & {\scriptsize{}0.03} & {\scriptsize{}0.03} & {\scriptsize{}0.03} & {\scriptsize{}0.03} & {\scriptsize{}0.03} & {\scriptsize{}0.03} & {\scriptsize{}0.02} & {\scriptsize{}0.03}\tabularnewline
\midrule
\multirow{3}{*}{$\eta_{1}$} & {\tiny{}Mean RelBias} & {\scriptsize{}-0.12} & {\scriptsize{}-0.02} & {\scriptsize{}0.09} & {\scriptsize{}-0.08} & {\scriptsize{}-0.07} & {\scriptsize{}-0.01} & {\scriptsize{}0.05} & {\scriptsize{}0.03} & {\scriptsize{}0.08} & {\scriptsize{}0.02} & {\scriptsize{}0.04} & {\scriptsize{}0.06} & {\scriptsize{}0.06} & {\scriptsize{}0.04}\tabularnewline
 & {\tiny{}Coverage} & {\scriptsize{}0.97} & {\scriptsize{}1.0} & {\scriptsize{}1.0} & {\scriptsize{}1.0} & {\scriptsize{}1.0} & {\scriptsize{}0.9} & {\scriptsize{}0.88} & {\scriptsize{}1.0} & {\scriptsize{}0.79} & {\scriptsize{}0.84} & {\scriptsize{}0.96} & {\scriptsize{}0.79} & {\scriptsize{}0.87} & {\scriptsize{}0.84}\tabularnewline
 & {\tiny{}Mean RelWidth} & {\scriptsize{}11.07} & {\scriptsize{}5.63} & {\scriptsize{}6.69} & {\scriptsize{}7.11} & {\scriptsize{}5.49} & {\scriptsize{}5.27} & {\scriptsize{}4.45} & {\scriptsize{}4.04} & {\scriptsize{}3.31} & {\scriptsize{}3.69} & {\scriptsize{}3.82} & {\scriptsize{}3.61} & {\scriptsize{}3.37} & {\scriptsize{}3.26}\tabularnewline
\midrule
\multirow{3}{*}{$\eta_{2}$} & {\tiny{}Mean RelBias} & {\scriptsize{}-0.06} & {\scriptsize{}0.03} & {\scriptsize{}0.1} & {\scriptsize{}-0.06} & {\scriptsize{}-0.05} & {\scriptsize{}-0.05} & {\scriptsize{}0.02} & {\scriptsize{}0.04} & {\scriptsize{}0.07} & {\scriptsize{}-0.01} & {\scriptsize{}0.02} & {\scriptsize{}0.04} & {\scriptsize{}0.04} & {\scriptsize{}0.02}\tabularnewline
 & {\tiny{}Coverage} & {\scriptsize{}0.94} & {\scriptsize{}1.0} & {\scriptsize{}1.0} & {\scriptsize{}0.96} & {\scriptsize{}1.0} & {\scriptsize{}0.9} & {\scriptsize{}0.93} & {\scriptsize{}0.97} & {\scriptsize{}0.83} & {\scriptsize{}0.84} & {\scriptsize{}0.96} & {\scriptsize{}0.88} & {\scriptsize{}0.9} & {\scriptsize{}0.89}\tabularnewline
 & {\tiny{}Mean RelWidth} & {\scriptsize{}9.97} & {\scriptsize{}4.56} & {\scriptsize{}7.65} & {\scriptsize{}9.63} & {\scriptsize{}7.02} & {\scriptsize{}8.9} & {\scriptsize{}6.33} & {\scriptsize{}3.56} & {\scriptsize{}3.04} & {\scriptsize{}2.92} & {\scriptsize{}5.95} & {\scriptsize{}5.9} & {\scriptsize{}4.99} & {\scriptsize{}5.36}\tabularnewline
\midrule
\multirow{3}{*}{$\gamma_{1}$} & {\tiny{}Mean RelBias} & {\scriptsize{}-0.29} & {\scriptsize{}-0.3} & {\scriptsize{}-0.59} & {\scriptsize{}-0.28} & {\scriptsize{}-0.11} & {\scriptsize{}-0.15} & {\scriptsize{}-0.1} & {\scriptsize{}-0.24} & {\scriptsize{}-0.23} & {\scriptsize{}-0.4} & {\scriptsize{}-0.19} & {\scriptsize{}-0.07} & {\scriptsize{}-0.16} & {\scriptsize{}-0.16}\tabularnewline
 & {\tiny{}Coverage} & {\scriptsize{}0.91} & {\scriptsize{}0.97} & {\scriptsize{}0.89} & {\scriptsize{}0.96} & {\scriptsize{}0.97} & {\scriptsize{}0.9} & {\scriptsize{}0.9} & {\scriptsize{}0.97} & {\scriptsize{}0.81} & {\scriptsize{}0.91} & {\scriptsize{}0.92} & {\scriptsize{}0.86} & {\scriptsize{}0.88} & {\scriptsize{}0.85}\tabularnewline
 & {\tiny{}Mean RelWidth} & {\scriptsize{}6.98} & {\scriptsize{}6.33} & {\scriptsize{}6.07} & {\scriptsize{}5.81} & {\scriptsize{}5.47} & {\scriptsize{}4.95} & {\scriptsize{}4.94} & {\scriptsize{}4.67} & {\scriptsize{}4.58} & {\scriptsize{}5.69} & {\scriptsize{}4.77} & {\scriptsize{}4.41} & {\scriptsize{}4.18} & {\scriptsize{}4.18}\tabularnewline
\midrule
\multirow{3}{*}{$\gamma_{2}$} & {\tiny{}Mean RelBias} & {\scriptsize{}-0.2} & {\scriptsize{}-0.41} & {\scriptsize{}-0.02} & {\scriptsize{}-0.24} & {\scriptsize{}-0.2} & {\scriptsize{}-0.22} & {\scriptsize{}-0.37} & {\scriptsize{}-0.31} & {\scriptsize{}-0.3} & {\scriptsize{}-0.41} & {\scriptsize{}-0.25} & {\scriptsize{}-0.29} & {\scriptsize{}-0.27} & {\scriptsize{}-0.3}\tabularnewline
 & {\tiny{}Coverage} & {\scriptsize{}0.94} & {\scriptsize{}0.94} & {\scriptsize{}1.0} & {\scriptsize{}0.96} & {\scriptsize{}0.97} & {\scriptsize{}0.9} & {\scriptsize{}0.95} & {\scriptsize{}0.95} & {\scriptsize{}0.95} & {\scriptsize{}0.96} & {\scriptsize{}0.98} & {\scriptsize{}0.95} & {\scriptsize{}0.96} & {\scriptsize{}0.95}\tabularnewline
 & {\tiny{}Mean RelWidth} & {\scriptsize{}9.01} & {\scriptsize{}8.29} & {\scriptsize{}9.48} & {\scriptsize{}7.94} & {\scriptsize{}7.7} & {\scriptsize{}6.97} & {\scriptsize{}7.38} & {\scriptsize{}7.41} & {\scriptsize{}7.19} & {\scriptsize{}8.41} & {\scriptsize{}6.88} & {\scriptsize{}6.63} & {\scriptsize{}6.5} & {\scriptsize{}6.39}\tabularnewline
\midrule
\multirow{3}{*}{$\sigma_{b}$} & {\tiny{}Mean RelBias} & {\scriptsize{}-0.78} & {\scriptsize{}-0.68} & {\scriptsize{}-0.22} & {\scriptsize{}-0.61} & {\scriptsize{}-1.06} & {\scriptsize{}-0.73} & {\scriptsize{}-0.57} & {\scriptsize{}-0.71} & {\scriptsize{}-0.51} & {\scriptsize{}-0.35} & {\scriptsize{}-0.37} & {\scriptsize{}-0.64} & {\scriptsize{}-0.65} & {\scriptsize{}-0.52}\tabularnewline
 & {\tiny{}Coverage} & {\scriptsize{}1.0} & {\scriptsize{}1.0} & {\scriptsize{}1.0} & {\scriptsize{}1.0} & {\scriptsize{}1.0} & {\scriptsize{}1.0} & {\scriptsize{}1.0} & {\scriptsize{}1.0} & {\scriptsize{}1.0} & {\scriptsize{}1.0} & {\scriptsize{}1.0} & {\scriptsize{}1.0} & {\scriptsize{}1.0} & {\scriptsize{}1.0}\tabularnewline
 & {\tiny{}Mean RelWidth} & {\scriptsize{}0.12} & {\scriptsize{}0.11} & {\scriptsize{}0.08} & {\scriptsize{}0.1} & {\scriptsize{}0.1} & {\scriptsize{}0.1} & {\scriptsize{}0.1} & {\scriptsize{}0.09} & {\scriptsize{}0.08} & {\scriptsize{}0.08} & {\scriptsize{}0.08} & {\scriptsize{}0.09} & {\scriptsize{}0.08} & {\scriptsize{}0.08}\tabularnewline
\midrule
\multirow{3}{*}{$\sigma_{t}$} & {\tiny{}Mean RelBias} & {\scriptsize{}-0.03} & {\scriptsize{}0.01} & {\scriptsize{}0.19} & {\scriptsize{}0.07} & {\scriptsize{}0.25} & {\scriptsize{}0.24} & {\scriptsize{}0.26} & {\scriptsize{}0.21} & {\scriptsize{}0.2} & {\scriptsize{}0.22} & {\scriptsize{}0.24} & {\scriptsize{}0.29} & {\scriptsize{}0.28} & {\scriptsize{}0.28}\tabularnewline
 & {\tiny{}Coverage} & {\scriptsize{}1.0} & {\scriptsize{}1.0} & {\scriptsize{}1.0} & {\scriptsize{}1.0} & {\scriptsize{}1.0} & {\scriptsize{}1.0} & {\scriptsize{}1.0} & {\scriptsize{}1.0} & {\scriptsize{}1.0} & {\scriptsize{}0.93} & {\scriptsize{}0.92} & {\scriptsize{}0.88} & {\scriptsize{}0.85} & {\scriptsize{}0.85}\tabularnewline
 & {\tiny{}Mean RelWidth} & {\scriptsize{}0.08} & {\scriptsize{}0.07} & {\scriptsize{}0.05} & {\scriptsize{}0.06} & {\scriptsize{}0.06} & {\scriptsize{}0.05} & {\scriptsize{}0.05} & {\scriptsize{}0.05} & {\scriptsize{}0.04} & {\scriptsize{}0.04} & {\scriptsize{}0.04} & {\scriptsize{}0.04} & {\scriptsize{}0.03} & {\scriptsize{}0.03}\tabularnewline
\midrule
\multirow{3}{*}{$\mu_{1}$} & {\tiny{}Mean RelBias} & {\scriptsize{}-0.02} & {\scriptsize{}-0.02} & {\scriptsize{}-0.02} & {\scriptsize{}-0.03} & {\scriptsize{}-0.02} & {\scriptsize{}-0.03} & {\scriptsize{}-0.03} & {\scriptsize{}-0.02} & {\scriptsize{}-0.02} & {\scriptsize{}-0.01} & {\scriptsize{}-0.03} & {\scriptsize{}-0.02} & {\scriptsize{}-0.03} & {\scriptsize{}-0.02}\tabularnewline
 & {\tiny{}Coverage} & {\scriptsize{}0.97} & {\scriptsize{}0.97} & {\scriptsize{}1.0} & {\scriptsize{}0.98} & {\scriptsize{}0.97} & {\scriptsize{}0.98} & {\scriptsize{}0.98} & {\scriptsize{}1.0} & {\scriptsize{}0.98} & {\scriptsize{}0.98} & {\scriptsize{}0.98} & {\scriptsize{}0.98} & {\scriptsize{}0.98} & {\scriptsize{}1.0}\tabularnewline
 & {\tiny{}Mean RelWidth} & {\scriptsize{}137.09} & {\scriptsize{}110.91} & {\scriptsize{}93.11} & {\scriptsize{}120.49} & {\scriptsize{}108.56} & {\scriptsize{}117.0} & {\scriptsize{}109.34} & {\scriptsize{}92.34} & {\scriptsize{}88.81} & {\scriptsize{}92.87} & {\scriptsize{}93.94} & {\scriptsize{}106.56} & {\scriptsize{}99.12} & {\scriptsize{}98.0}\tabularnewline
\midrule 
\multirow{3}{*}{$\mu_{2}$} & {\tiny{}Mean RelBias} & {\scriptsize{}0.02} & {\scriptsize{}0.02} & {\scriptsize{}0.01} & {\scriptsize{}0.02} & {\scriptsize{}0.02} & {\scriptsize{}0.02} & {\scriptsize{}0.02} & {\scriptsize{}0.02} & {\scriptsize{}0.01} & {\scriptsize{}0.01} & {\scriptsize{}0.02} & {\scriptsize{}0.02} & {\scriptsize{}0.02} & {\scriptsize{}0.02}\tabularnewline
 & {\tiny{}Coverage} & {\scriptsize{}1.0} & {\scriptsize{}1.0} & {\scriptsize{}1.0} & {\scriptsize{}1.0} & {\scriptsize{}1.0} & {\scriptsize{}1.0} & {\scriptsize{}1.0} & {\scriptsize{}1.0} & {\scriptsize{}1.0} & {\scriptsize{}1.0} & {\scriptsize{}1.0} & {\scriptsize{}1.0} & {\scriptsize{}1.0} & {\scriptsize{}1.0}\tabularnewline
 & {\tiny{}Mean RelWidth} & {\scriptsize{}394.42} & {\scriptsize{}371.94} & {\scriptsize{}273.44} & {\scriptsize{}355.58} & {\scriptsize{}347.17} & {\scriptsize{}375.67} & {\scriptsize{}373.83} & {\scriptsize{}340.18} & {\scriptsize{}336.57} & {\scriptsize{}326.04} & {\scriptsize{}319.8} & {\scriptsize{}374.47} & {\scriptsize{}353.83} & {\scriptsize{}357.8}\tabularnewline
\bottomrule
\end{tabular}
}
\caption{\label{tab:3ReplicatesCoverages} Mean relative bias, coverage and mean relative width of the 95\% credible interval for each parameter (rows) and for all the designs (columns) when the number of replicates per observation time where fixed to 3 mice.  For each entry in the table, 60 different synthetic datasets were used.}
\end{table}

\begin{table}[ht]
\resizebox{\columnwidth}{!}{%
    \addtolength{\leftskip} {-2cm}
    \addtolength{\rightskip}{-2cm}
\begin{tabular}{cccccccccccccccc}
\toprule 
\begin{turn}{90}
{\scriptsize{}Parameter}
\end{turn} & \begin{turn}{90}
{\scriptsize{}7 replicates}
\end{turn} & \begin{turn}{90}
{\scriptsize{}0,6}
\end{turn} & \begin{turn}{90}
{\scriptsize{}0,0.5,6}
\end{turn} & \begin{turn}{90}
{\scriptsize{}0,0.5,1}
\end{turn} & \begin{turn}{90}
{\scriptsize{}0,3,6}
\end{turn} & \begin{turn}{90}
{\scriptsize{}0,1,6}
\end{turn} & \begin{turn}{90}
{\scriptsize{}0,2,4,6}
\end{turn} & \begin{turn}{90}
{\scriptsize{}0,1,2,6}
\end{turn} & \begin{turn}{90}
{\scriptsize{}0,0.5,1,6}
\end{turn} & \begin{turn}{90}
{\scriptsize{}0,0.5,1,2,6}
\end{turn} & \begin{turn}{90}
{\scriptsize{}0,0.5,1,2,3}
\end{turn} & \begin{turn}{90}
{\scriptsize{}0,1,2,3,6}
\end{turn} & \begin{turn}{90}
{\scriptsize{}0,1,2,4,6}
\end{turn} & \begin{turn}{90}
{\scriptsize{}0,1,2,3,4,6}
\end{turn} & \begin{turn}{90}
{\scriptsize{}0,1,2,3,4,5,6}
\end{turn}\tabularnewline
\midrule 
\multirow{3}{*}{$p_{0}$} & {\tiny{}Mean RelBias} & {\scriptsize{}0.01} & {\scriptsize{}0.01} & {\scriptsize{}-0.01} & {\scriptsize{}0.0} & {\scriptsize{}0.01} & {\scriptsize{}-0.0} & {\scriptsize{}0.0} & {\scriptsize{}0.0} & {\scriptsize{}0.0} & {\scriptsize{}-0.0} & {\scriptsize{}0.0} & {\scriptsize{}0.0} & {\scriptsize{}0.0} & {\scriptsize{}-0.0}\tabularnewline
 & {\tiny{}Coverage} & {\scriptsize{}0.97} & {\scriptsize{}0.97} & {\scriptsize{}1.0} & {\scriptsize{}0.98} & {\scriptsize{}1.0} & {\scriptsize{}0.96} & {\scriptsize{}0.97} & {\scriptsize{}1.0} & {\scriptsize{}1.0} & {\scriptsize{}1.0} & {\scriptsize{}0.94} & {\scriptsize{}0.96} & {\scriptsize{}0.97} & {\scriptsize{}0.97}\tabularnewline
 & {\tiny{}Mean RelWidth} & {\scriptsize{}0.04} & {\scriptsize{}0.04} & {\scriptsize{}0.04} & {\scriptsize{}0.04} & {\scriptsize{}0.03} & {\scriptsize{}0.03} & {\scriptsize{}0.02} & {\scriptsize{}0.02} & {\scriptsize{}0.02} & {\scriptsize{}0.03} & {\scriptsize{}0.02} & {\scriptsize{}0.02} & {\scriptsize{}0.02} & {\scriptsize{}0.02}\tabularnewline
\midrule
\multirow{3}{*}{$\eta_{1}$} & {\tiny{}Mean RelBias} & {\scriptsize{}0.07} & {\scriptsize{}0.02} & {\scriptsize{}0.05} & {\scriptsize{}0.06} & {\scriptsize{}0.03} & {\scriptsize{}0.1} & {\scriptsize{}0.07} & {\scriptsize{}0.04} & {\scriptsize{}0.06} & {\scriptsize{}0.04} & {\scriptsize{}0.07} & {\scriptsize{}0.07} & {\scriptsize{}0.06} & {\scriptsize{}0.1}\tabularnewline
 & {\tiny{}Coverage} & {\scriptsize{}0.97} & {\scriptsize{}0.97} & {\scriptsize{}1.0} & {\scriptsize{}0.94} & {\scriptsize{}0.97} & {\scriptsize{}0.93} & {\scriptsize{}0.97} & {\scriptsize{}0.97} & {\scriptsize{}0.94} & {\scriptsize{}1.0} & {\scriptsize{}0.97} & {\scriptsize{}0.96} & {\scriptsize{}1.0} & {\scriptsize{}0.94}\tabularnewline
 & {\tiny{}Mean RelWidth} & {\scriptsize{}10.31} & {\scriptsize{}5.05} & {\scriptsize{}4.63} & {\scriptsize{}5.41} & {\scriptsize{}3.92} & {\scriptsize{}4.36} & {\scriptsize{}2.94} & {\scriptsize{}2.99} & {\scriptsize{}2.19} & {\scriptsize{}2.6} & {\scriptsize{}2.62} & {\scriptsize{}3.07} & {\scriptsize{}3.26} & {\scriptsize{}3.36}\tabularnewline
\midrule
\multirow{3}{*}{$\eta_{2}$} & {\tiny{}Mean RelBias} & {\scriptsize{}0.08} & {\scriptsize{}0.02} & {\scriptsize{}-0.02} & {\scriptsize{}0.02} & {\scriptsize{}-0.02} & {\scriptsize{}0.06} & {\scriptsize{}0.03} & {\scriptsize{}0.02} & {\scriptsize{}0.03} & {\scriptsize{}0.01} & {\scriptsize{}0.02} & {\scriptsize{}0.04} & {\scriptsize{}0.02} & {\scriptsize{}0.05}\tabularnewline
 & {\tiny{}Coverage} & {\scriptsize{}0.95} & {\scriptsize{}0.97} & {\scriptsize{}1.0} & {\scriptsize{}0.92} & {\scriptsize{}0.97} & {\scriptsize{}0.95} & {\scriptsize{}0.97} & {\scriptsize{}1.0} & {\scriptsize{}1.0} & {\scriptsize{}1.0} & {\scriptsize{}0.97} & {\scriptsize{}0.96} & {\scriptsize{}1.0} & {\scriptsize{}0.94}\tabularnewline
 & {\tiny{}Mean RelWidth} & {\scriptsize{}7.34} & {\scriptsize{}4.35} & {\scriptsize{}5.09} & {\scriptsize{}8.0} & {\scriptsize{}5.19} & {\scriptsize{}6.79} & {\scriptsize{}4.12} & {\scriptsize{}3.08} & {\scriptsize{}2.48} & {\scriptsize{}3.08} & {\scriptsize{}3.67} & {\scriptsize{}4.39} & {\scriptsize{}4.7} & {\scriptsize{}4.38}\tabularnewline
\midrule
\multirow{3}{*}{$\gamma_{1}$} & {\tiny{}Mean RelBias} & {\scriptsize{}-0.07} & {\scriptsize{}0.03} & {\scriptsize{}-0.46} & {\scriptsize{}-0.05} & {\scriptsize{}0.05} & {\scriptsize{}-0.05} & {\scriptsize{}-0.0} & {\scriptsize{}-0.0} & {\scriptsize{}0.02} & {\scriptsize{}-0.36} & {\scriptsize{}0.04} & {\scriptsize{}0.01} & {\scriptsize{}-0.01} & {\scriptsize{}-0.08}\tabularnewline
 & {\tiny{}Coverage} & {\scriptsize{}0.92} & {\scriptsize{}0.95} & {\scriptsize{}0.89} & {\scriptsize{}0.96} & {\scriptsize{}0.91} & {\scriptsize{}0.96} & {\scriptsize{}0.92} & {\scriptsize{}0.97} & {\scriptsize{}0.98} & {\scriptsize{}0.9} & {\scriptsize{}1.0} & {\scriptsize{}0.98} & {\scriptsize{}0.97} & {\scriptsize{}1.0}\tabularnewline
 & {\tiny{}Mean RelWidth} & {\scriptsize{}5.88} & {\scriptsize{}4.65} & {\scriptsize{}6.14} & {\scriptsize{}4.87} & {\scriptsize{}4.35} & {\scriptsize{}4.29} & {\scriptsize{}3.48} & {\scriptsize{}3.47} & {\scriptsize{}3.18} & {\scriptsize{}4.83} & {\scriptsize{}3.0} & {\scriptsize{}3.22} & {\scriptsize{}3.27} & {\scriptsize{}3.38}\tabularnewline
\midrule
\multirow{3}{*}{$\gamma_{2}$} & {\tiny{}Mean RelBias} & {\scriptsize{}-0.3} & {\scriptsize{}-0.27} & {\scriptsize{}-0.36} & {\scriptsize{}-0.41} & {\scriptsize{}-0.45} & {\scriptsize{}-0.37} & {\scriptsize{}-0.26} & {\scriptsize{}-0.24} & {\scriptsize{}-0.27} & {\scriptsize{}-0.23} & {\scriptsize{}-0.48} & {\scriptsize{}-0.31} & {\scriptsize{}-0.33} & {\scriptsize{}-0.47}\tabularnewline
 & {\tiny{}Coverage} & {\scriptsize{}0.92} & {\scriptsize{}0.97} & {\scriptsize{}0.93} & {\scriptsize{}0.96} & {\scriptsize{}0.88} & {\scriptsize{}0.93} & {\scriptsize{}0.97} & {\scriptsize{}0.94} & {\scriptsize{}0.94} & {\scriptsize{}0.96} & {\scriptsize{}0.88} & {\scriptsize{}0.92} & {\scriptsize{}0.95} & {\scriptsize{}0.89}\tabularnewline
 & {\tiny{}Mean RelWidth} & {\scriptsize{}9.64} & {\scriptsize{}8.33} & {\scriptsize{}7.95} & {\scriptsize{}7.86} & {\scriptsize{}7.62} & {\scriptsize{}6.8} & {\scriptsize{}5.69} & {\scriptsize{}6.22} & {\scriptsize{}5.82} & {\scriptsize{}6.84} & {\scriptsize{}6.2} & {\scriptsize{}5.99} & {\scriptsize{}6.37} & {\scriptsize{}6.1}\tabularnewline
\midrule
\multirow{3}{*}{$\sigma_{b}$} & {\tiny{}Mean RelBias} & {\scriptsize{}-0.74} & {\scriptsize{}-0.51} & {\scriptsize{}-0.39} & {\scriptsize{}-0.63} & {\scriptsize{}-0.99} & {\scriptsize{}-0.47} & {\scriptsize{}-0.7} & {\scriptsize{}-0.64} & {\scriptsize{}-0.59} & {\scriptsize{}-0.41} & {\scriptsize{}-0.41} & {\scriptsize{}-0.53} & {\scriptsize{}-0.54} & {\scriptsize{}-0.31}\tabularnewline
 & {\tiny{}Coverage} & {\scriptsize{}1.0} & {\scriptsize{}1.0} & {\scriptsize{}1.0} & {\scriptsize{}1.0} & {\scriptsize{}0.97} & {\scriptsize{}1.0} & {\scriptsize{}1.0} & {\scriptsize{}1.0} & {\scriptsize{}1.0} & {\scriptsize{}1.0} & {\scriptsize{}1.0} & {\scriptsize{}1.0} & {\scriptsize{}1.0} & {\scriptsize{}1.0}\tabularnewline
 & {\tiny{}Mean RelWidth} & {\scriptsize{}0.1} & {\scriptsize{}0.09} & {\scriptsize{}0.08} & {\scriptsize{}0.08} & {\scriptsize{}0.08} & {\scriptsize{}0.07} & {\scriptsize{}0.06} & {\scriptsize{}0.08} & {\scriptsize{}0.07} & {\scriptsize{}0.06} & {\scriptsize{}0.07} & {\scriptsize{}0.06} & {\scriptsize{}0.07} & {\scriptsize{}0.06}\tabularnewline
\midrule
\multirow{3}{*}{$\sigma_{t}$} & {\tiny{}Mean RelBias} & {\scriptsize{}-0.02} & {\scriptsize{}-0.06} & {\scriptsize{}0.12} & {\scriptsize{}0.02} & {\scriptsize{}0.2} & {\scriptsize{}0.07} & {\scriptsize{}0.19} & {\scriptsize{}0.13} & {\scriptsize{}0.13} & {\scriptsize{}0.13} & {\scriptsize{}0.15} & {\scriptsize{}0.18} & {\scriptsize{}0.17} & {\scriptsize{}0.07}\tabularnewline
 & {\tiny{}Coverage} & {\scriptsize{}1.0} & {\scriptsize{}1.0} & {\scriptsize{}1.0} & {\scriptsize{}1.0} & {\scriptsize{}0.97} & {\scriptsize{}1.0} & {\scriptsize{}1.0} & {\scriptsize{}1.0} & {\scriptsize{}1.0} & {\scriptsize{}1.0} & {\scriptsize{}1.0} & {\scriptsize{}0.98} & {\scriptsize{}0.95} & {\scriptsize{}1.0}\tabularnewline
 & {\tiny{}Mean RelWidth} & {\scriptsize{}0.06} & {\scriptsize{}0.05} & {\scriptsize{}0.04} & {\scriptsize{}0.04} & {\scriptsize{}0.04} & {\scriptsize{}0.03} & {\scriptsize{}0.03} & {\scriptsize{}0.04} & {\scriptsize{}0.03} & {\scriptsize{}0.03} & {\scriptsize{}0.03} & {\scriptsize{}0.02} & {\scriptsize{}0.02} & {\scriptsize{}0.02}\tabularnewline
\midrule
\multirow{3}{*}{$\mu_{1}$} & {\tiny{}Mean RelBias} & {\scriptsize{}-0.04} & {\scriptsize{}-0.04} & {\scriptsize{}-0.04} & {\scriptsize{}-0.04} & {\scriptsize{}-0.04} & {\scriptsize{}-0.04} & {\scriptsize{}-0.03} & {\scriptsize{}-0.05} & {\scriptsize{}-0.04} & {\scriptsize{}-0.05} & {\scriptsize{}-0.03} & {\scriptsize{}-0.03} & {\scriptsize{}-0.04} & {\scriptsize{}-0.03}\tabularnewline
 & {\tiny{}Coverage} & {\scriptsize{}0.95} & {\scriptsize{}0.9} & {\scriptsize{}0.85} & {\scriptsize{}0.96} & {\scriptsize{}0.91} & {\scriptsize{}0.95} & {\scriptsize{}0.97} & {\scriptsize{}0.79} & {\scriptsize{}0.73} & {\scriptsize{}0.62} & {\scriptsize{}0.91} & {\scriptsize{}0.92} & {\scriptsize{}0.9} & {\scriptsize{}0.94}\tabularnewline
 & {\tiny{}Mean RelWidth} & {\scriptsize{}96.41} & {\scriptsize{}83.51} & {\scriptsize{}85.52} & {\scriptsize{}89.44} & {\scriptsize{}82.03} & {\scriptsize{}83.48} & {\scriptsize{}66.3} & {\scriptsize{}73.97} & {\scriptsize{}63.96} & {\scriptsize{}72.4} & {\scriptsize{}73.12} & {\scriptsize{}70.75} & {\scriptsize{}73.08} & {\scriptsize{}72.03}\tabularnewline
\midrule 
\multirow{3}{*}{$\mu_{2}$} & {\tiny{}Mean RelBias} & {\scriptsize{}0.05} & {\scriptsize{}0.05} & {\scriptsize{}0.05} & {\scriptsize{}0.05} & {\scriptsize{}0.05} & {\scriptsize{}0.05} & {\scriptsize{}0.04} & {\scriptsize{}0.05} & {\scriptsize{}0.05} & {\scriptsize{}0.05} & {\scriptsize{}0.05} & {\scriptsize{}0.05} & {\scriptsize{}0.05} & {\scriptsize{}0.05}\tabularnewline
 & {\tiny{}Coverage} & {\scriptsize{}0.92} & {\scriptsize{}0.95} & {\scriptsize{}0.96} & {\scriptsize{}0.92} & {\scriptsize{}0.88} & {\scriptsize{}0.93} & {\scriptsize{}0.95} & {\scriptsize{}0.97} & {\scriptsize{}0.94} & {\scriptsize{}0.92} & {\scriptsize{}0.91} & {\scriptsize{}0.88} & {\scriptsize{}0.87} & {\scriptsize{}0.86}\tabularnewline
 & {\tiny{}Mean RelWidth} & {\scriptsize{}255.16} & {\scriptsize{}250.85} & {\scriptsize{}233.93} & {\scriptsize{}255.83} & {\scriptsize{}253.79} & {\scriptsize{}249.89} & {\scriptsize{}209.19} & {\scriptsize{}244.21} & {\scriptsize{}227.82} & {\scriptsize{}230.94} & {\scriptsize{}238.74} & {\scriptsize{}231.04} & {\scriptsize{}230.26} & {\scriptsize{}226.08}\tabularnewline
\bottomrule
\end{tabular}
}
\caption{\label{tab:7ReplicatesCoverages} Mean relative bias, coverage and mean relative width of the 95\% credible interval for each parameter (rows) and for all the designs (columns) when the number of replicates per observation time is restricted to 7 mice.  For each entry in the table, 60 different synthetic datasets were used.}
\end{table}

\begin{figure}[!b]
    \centering
    \includegraphics[width=0.95\textwidth]{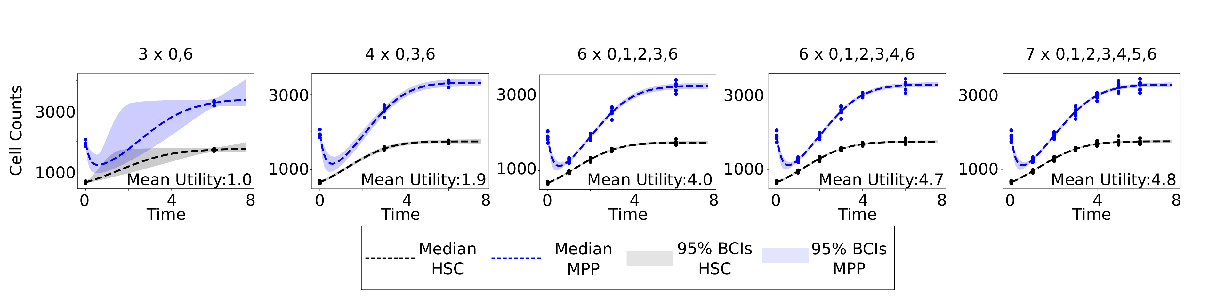}
    \caption{\textbf{Comparing information gain and goodness of fit for multiple designs} We show the ODE solutions posterior distributions for HSC and MPPs for multiple designs. 
    We selected designs with increasing value of the mean utility from Fig. \ref{fig:MeanUtilitiesHeatMap}.
    The median ODE solution (dashed), the 95\% Bayesian credible intervals (bands) and the mean utility values are shown.}
  \label{fig:comparing_utilities}
  \vspace{-12pt}
\end{figure}

\begin{figure}[ht]
    \centering
    \includegraphics[width=0.95\textwidth]{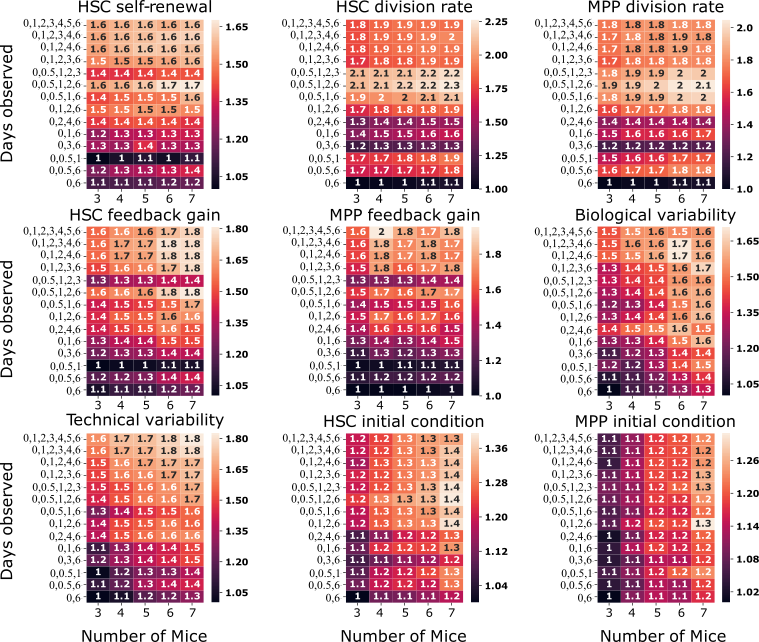}
    \caption{\textbf{Mean utility values for all the model parameters.} The heat maps show the mean utility value for the labeled parameter normalized by the lowest mean utility. Each parameter utility heat map is scaled differently.}
  \label{fig:heat_map_parameter_utils_all}
  \vspace{-8pt}
\end{figure}

\newpage
\begin{figure}[ht]
    \centering
    \includegraphics[width=0.8\textwidth]{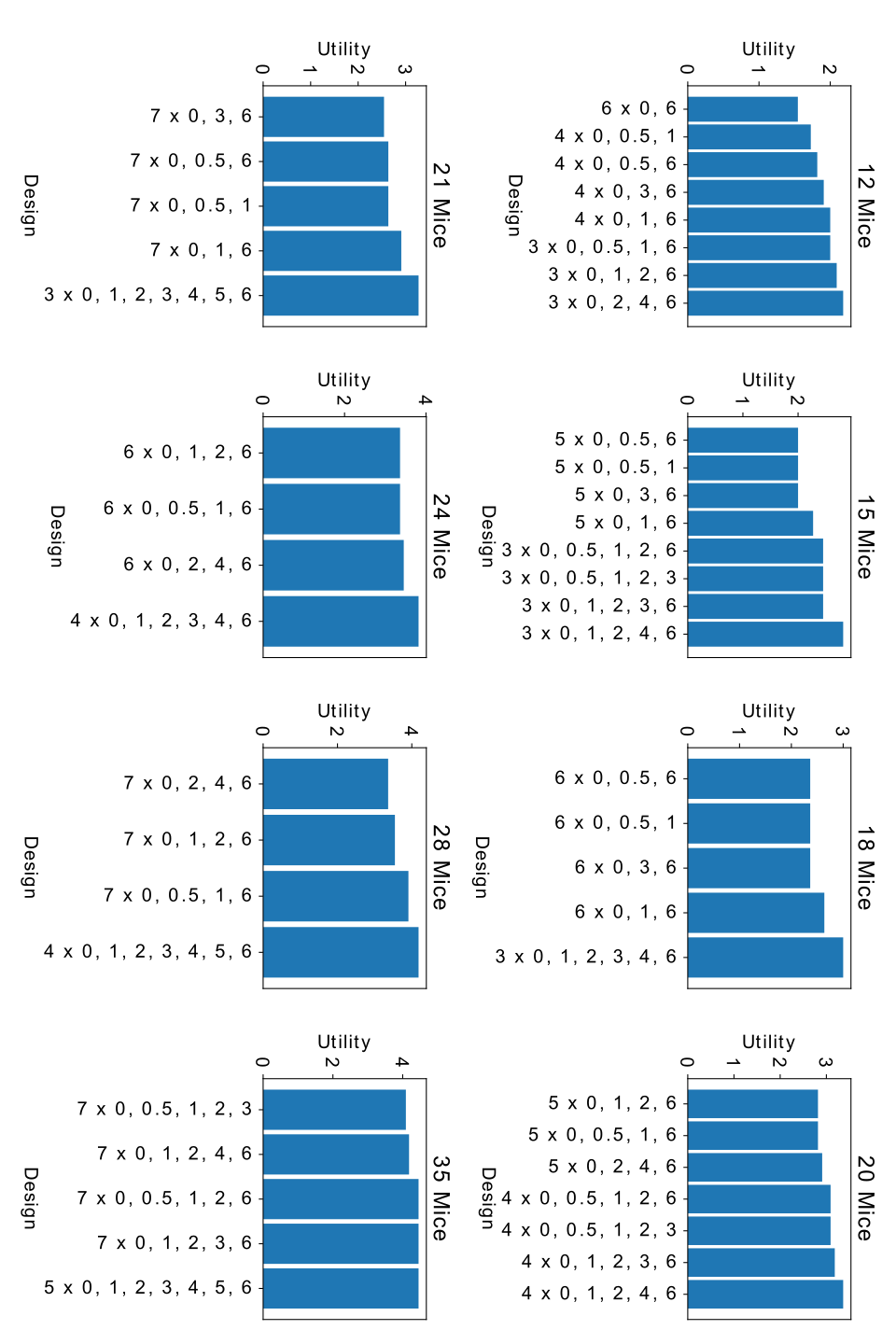}
    \caption{\textbf{Comparing designs with a fixed number of mice.} The mean utility values of the different designs are compared when the same number of mice is used. The utility value is scaled as in the overall utility from Fig. \ref{fig:MeanUtilitiesHeatMap}}
  \label{fig:fixed_mice_comparison}
  \vspace{-8pt}
\end{figure}

\newpage
\begin{figure}[!b]
    \centering
    \includegraphics[width=0.95\textwidth]{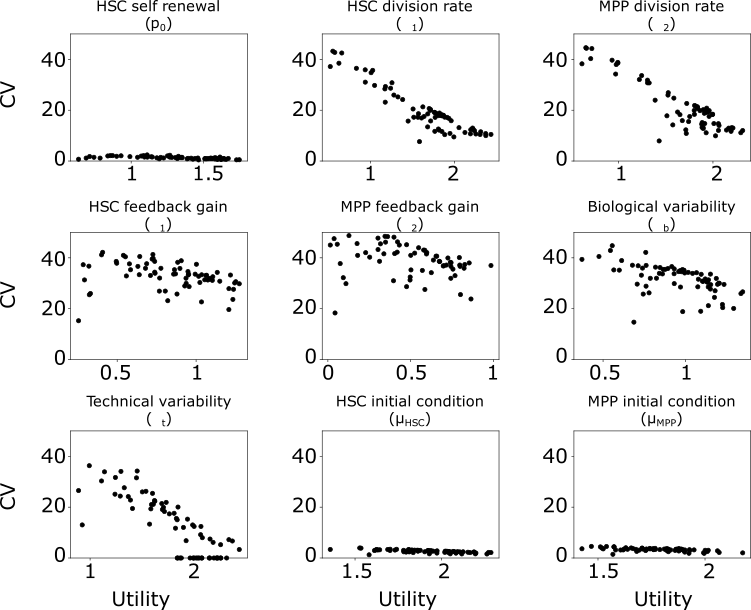}
    \caption{\textbf{Coefficient of variation and mean utility values} The mean coefficient of variation (CV) for all designs are obtained and plotted against the parameter utility value. The utility values are raw values and are not normalized like the utilities in the heat maps.}
  \label{fig:cv_vs_utility}
  \vspace{-12pt}
\end{figure}

\begin{figure}[ht]
    \centering
    \includegraphics[width=0.95\textwidth]{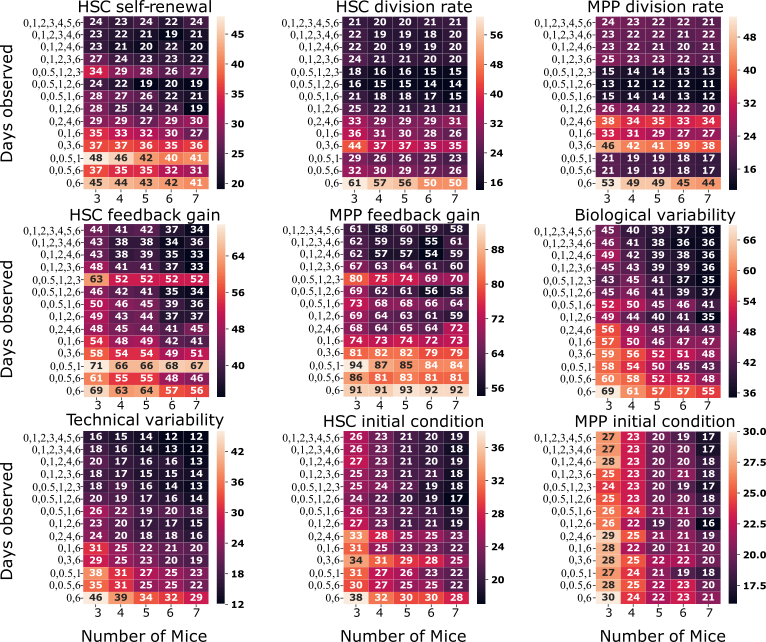}
    \caption{\textbf{Prior vs posterior mean percent overlap for individual parameters} We calculated the mean percent overlap between the prior and posterior distributions as a complementary measure to the mean utility value for information gain.}
    \label{fig:percent_overlap}
  \vspace{-12pt}
\end{figure}

\begin{figure}[ht]
    \centering
    \includegraphics[width=0.8\textwidth]{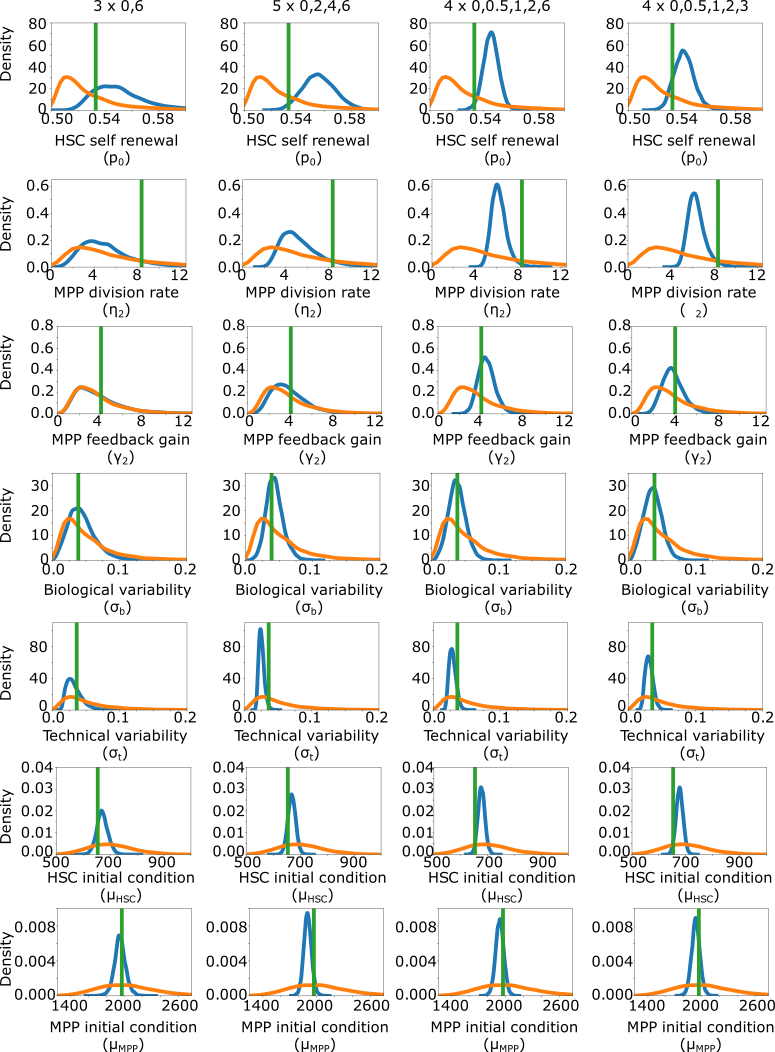}
    \caption{\textbf{Prior vs posterior distributions for sampled designs} We show the remaining prior and posterior distributions for the different parameters (rows) and designs (columns) that were not shown in Fig. \ref{fig:Gam1Eta1Utilities} in the main text.}
  \label{fig:pvp_param_util_all}
  \vspace{-12pt}
\end{figure}

\end{document}